\pdfoutput=1

\documentclass[11pt,showonlyrefs]{arxiv-article}
\usepackage{fixltx2e}

\bibliography{References,T-duality,New3dDualities}

\usepackage{Definitions}

\usepackage{sidecap}
\sidecaptionvpos{figure}{c}

\newcommand*{\splitcell}[3]{\begin{tabular}{#1}#2\\#3\end{tabular}}

\RequirePackage{hyperref}

\allowdisplaybreaks[2]

\newcommand{\OfficialTitle}{
Monopole Quivers\\and new 3D N=2 dualities
}
\title{\vspace{2cm}
  {\color{Thoughtless}\Huge\textbf{\dosserif\OfficialTitle}}
}

\hypersetup{pdfauthor={Antonio Amariti and Domenico Orlando and Susanne Reffert},pdftitle={\OfficialTitle}}

\author{%
  \begin{minipage}{.8\linewidth}
    \vspace{1cm}
    \begin{center} \dosserif
      {\small 
        \textbf{Antonio~Amariti},
         \textbf{Domenico~Orlando} and
        \textbf{Susanne~Reffert}}
    \end{center}
    \vspace{1cm}
    \authorBlock{}{Albert Einstein Center for Fundamental Physics\\
      Institute for Theoretical Physics\\
      University of Bern,\\
      Sidlerstrasse 5, \textsc{ch}-3012 Bern, Switzerland}
  \end{minipage}
}

\date{} 

\def\O#1{\ensuremath{\mathrm{O}#1}}
\begin{document}

\setstretch{1.15}

\numberwithin{equation}{section}

\begin{titlepage}

  \newgeometry{top=23.1mm,bottom=46.1mm,left=34.6mm,right=34.6mm}

  \maketitle

  \thispagestyle{empty}

  \vfill\dosserif
  
  \abstract{\normalfont \noindent

    We present a new family of dualities for three-dimensional gauge theories, motivated by the brane realization of the reduction of four-dimensional dualities on a circle.
    This family can be understood as a generalization of Aharony duality to quiver gauge theories whose nodes interact via monopole terms in the superpotential.
    We refer to this family of theories as \emph{monopole quivers}.
    We corroborate the new dualities by checking the equivalence of the three-sphere partition functions, obtained from the standard circle reduction of the four-dimensional superconformal index.
    As a special case, we recover some dualities recently discussed in the literature.}

\vfill

\end{titlepage}

\restoregeometry

\tableofcontents

\section{Introduction}

The description of the low-energy dynamics of \ac{uv} free \acp{qft} often requires the use of non-perturbative techniques.
A different paradigm consists in describing the theory in terms of dual degrees of freedom, hopefully weakly coupled at low energy scales.
Large classes of such dualities have been worked out in the case of supersymmetric \acp{qft}, mostly thanks to the power of holomorphy.
When considering four-dimensional theories with the minimal amount of supersymmetry, a rich duality web has been obtained, based on extensions and deformations of Seiberg duality~\cite{Seiberg:1994pq}.
It has been quickly realized that similar dualities exist for three-dimensional $\mathcal{N}=2$ theories~\cite{Aharony:1997bx,Aharony:1997gp,Karch:1997ux,Giveon:2008zn,Benini:2011mf,Benini:2017dud}.
The similarity has been explained in~\cite{Aharony:2013dha}, where three-dimensional dualities have been derived from a circle compactification of four-dimensional Seiberg dualities.

The web of three-dimensional dualities turns out to be richer than the one in four dimensions, mostly due to the presence of a Coulomb branch for three-dimensional theories with four supercharges.
This Coulomb branch is described in terms of monopole operators, associated to chiral fields consisting of a combination of a real adjoint scalar in the vector multiplet and of the dual photon.
These operators have been used to enlarge the spectrum of the three-dimensional dualities.

In general, new dualities emerge if the compactification limit is taken while 
performing real mass flows and Higgsing the gauge group in non-trivial vacua. A complete classification is still lacking to date and a full understanding of the behavior of the monopole operators in a circle compactification is a necessary step in this direction.

The reduction of four-dimensional dualities to three dimensions simplifies when the gauge theories are engineered in a setup of intersecting branes~\cite{Amariti:2015yea}.
The circle reduction corresponds to a T-duality from the \tIIA description of the four-dimensional theory to the \tIIB description of the three-dimensional theory.
The power of this picture stems from the fact that the monopole operators have a straightforward realization in terms of Euclidean \D1-branes.
Dual configurations in presence of non-trivial vacuum structures can be directly obtained via a \ac{hw} transition.

In this paper, we show that a very rich spectrum of new three-dimensional $\mathcal{N}=2$ dualities emerges from reducing four-dimensional dualities in the brane setup and considering non-trivial vacuum configurations.
In field theory, they correspond to products of three-dimensional \ac{sqcd}-like sectors interacting through \ac{ahw} superpotentials.
They can be represented as generalized quiver gauge theories, where the gauge nodes interact through monopole operators.
These monopole operators, responsible for the coupling of the various gauge nodes, reconstruct the original \ac{kk} monopole of the circle compactification.

In terms of the brane set-up, we will be using the same technique of the circle compactification as in~\cite{Amariti:2015yea}
and consider vacuum configurations obtained by spreading several stacks of \D3 branes on the circle, attaching them to \D5 branes.
The resulting theory is thus akin to a quiver gauge theory in which each of the stacks of \D3s corresponds to a node in the quiver. We will however take the circle radius to be very large, in which case the bifundamental fields, corresponding to fundamental strings stretched between the stacks vanish, while the monopoles, corresponding as before to \D1 branes stretched between the stacks of \D3s remain.
The stacks of \D3s thus only interact via the monopoles, which is why we coin the term \emph{monopole quiver} for this brane construction.

Using this set-up as a starting point, we can perform a variety of duality transformations via \ac{hw} transitions just as in~\cite{Amariti:2015yea,Amariti:2015mva}, resulting in new dualities.

\medskip
The plan of this paper is as follows. In Section~\ref{sec:review}, we review the brane construction introduced in~\cite{Amariti:2015yea} on which the dualities are realized as \ac{hw} moves.
In Section~\ref{sec:newfrom4D}, we derive a new duality between two monopole quivers with unitary gauge groups
by reducing four-dimensional Seiberg duality on $S^1$ and by performing a large real mass and Higgs flow to recover the three-dimensional limit.
In Section~\ref{sec:knownD} we discuss how to recover the dualities described in~\cite{Benini:2017dud} via our brane set-up. 
The generalization of the picture to monopole quivers with $Sp$ gauge groups is discussed in Section~\ref{sec:general}. 
In Sec.~\ref{sec:concl}, we end with concluding remarks and an outlook.
In appendix~\ref{app:generalduality}, we discuss the three-dimensional limit for the monopole quiver duality with a generic amount of unitary gauge groups, 
showing the matching of the partition function between the dual phases after the real mass and Higgs flow.
Appendix~\ref{appPF} is a review of the reduction of the four-dimensional superconformal index to the three-dimensional partition 
function on a squashed three-sphere, serving also to fix notation and conventions.

\section{The brane set-up on the circle}\label{sec:review}

\paragraph{Circle compactification.}
Let us first review the circle compactification as introduced in~\cite{Amariti:2015yea}.
We start with the brane set-up for $\mathcal{N}=1$ $U(N)$ \ac{sym} on $\mathbb{R}^3 \times S^1$,
denoting by  $R_3$ the radius of $S^1$.
It can be arrived at by a T--duality from a stack of $N$ \D4 branes suspended between an \NS5 and an \NS5' brane at distance \(\ell_6 \) in the \(x^6 \) direction.
Performing the T-duality along the compact $x^3$ direction, the \D4s turn into \D3s, while the \NS5 branes remain unchanged. The resulting configuration is summarized in Table~\ref{tab:sqcd-no-flavors}.

\begin{table}
  \centering
  \begin{tabular}{lCCCCCCCCCC}
    \toprule
        & 0 & 1 & 2 & 3 & 4 & 5 & 6 & 7 & 8 & 9 \\
    \midrule
    NS  & \times & \times & \times & \times & \times & \times &   &   &       \\
    NS' & \times & \times & \times & \times &   &   &   &   & \times & \times \\
    \D3 & \times & \times & \times &   &   &   & \times &   &       \\ \bottomrule
  \end{tabular}
  \caption{Brane configuration for \(\mathcal{N}=1\) \ac{sym} on $\mathbb{R}^3 \times S^1$. The \D3--branes are suspended between the two \NS5s.}
  \label{tab:sqcd-no-flavors}
\end{table}

The corresponding gauge theory has $N$ isolated vacua, corresponding to stable supersymmetric configurations of the brane system.
There is a repulsive force between the \D3s and, 
in a stable configuration, the \D3 branes are distributed along the circle direction $x^3$ at equal distances.
All moduli are lifted as the \D3s cannot move freely. %

\begin{figure}
  \begin{center}
    \begin{footnotesize}
      \begin{tikzpicture}
  \begin{scope}[shift={(-3,-2)}]
    \begin{footnotesize}
      \draw[->] (-0.2,0) arc (-140:135:.2 and .4);
      \draw (0.1,.9) node[]{\(x_3\)};
      \draw[->] (-0.4,.25) -- (.6,.25);
      \draw (0.6,.45) node[]{\(x_6\)};
    \end{footnotesize}
  \end{scope}

  \begin{scope}[thick]
    \begin{footnotesize}
      \draw[fill=black!10,black!10] (0,-.5) rectangle (7,.5); 
      \draw (0,-.5) -- (7,-.5);
      \draw (0,.5) -- (7,.5);
      \draw (.5,1.5) -- (6.7, 1.5);
      \draw (.5,-1.5) -- (6.7, -1.5);

      \draw[fill=white] (0,0) ellipse (1 and 2);
      \draw[dashed] (6,0) ellipse (1 and 2);

      \draw (0,2) -- (6,2);
      \draw (0,-2) -- (6,-2);

      \draw (-.8,-2) node[]{\NS5};
      \draw (6.7,-2) node[]{\NS5'};
      \draw (3,-1) node[]{\D3};         
      \draw (4,0) node[]{\D1};

    \end{footnotesize}
  \end{scope}
\end{tikzpicture}
    \end{footnotesize}
  \end{center}
  \caption{\D1--branes (in grey) stretched between \D3--branes and \NS5 branes.}
  \label{fig:D1-between-D3}
\end{figure}
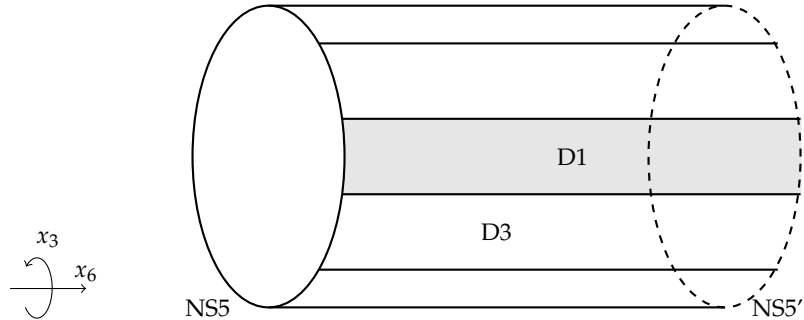

Let us discuss a moment the origin of this repulsive force.
From the field theory point of view in three dimensions, instantons induce a non-perturbative superpotential.
The instantons are represented in the brane picture by Euclidean \D1s stretched between each pair of \D3 branes along $x^3$ and in $x^6$ and the \NS{} and \NS' branes (see Figure~\ref{fig:D1-between-D3}).

The \D1--branes give rise to the superpotential~\cite{deBoer:1996ck}
\begin{equation}
  \label{WD1}
  W = \sum_i e^{-S_{{\D1}_i}} = \sum_{i=1}^{N-1} \exp[\Sigma_{i+1}-\Sigma_{i}],
\end{equation}
with $\Sigma_i = \sigma_i/e_3^2 + i \phi_i $, where the scalar $\sigma_i$ parameterizes the position of the \(i\)--th \D3, $\phi$ is the dual photon, and $e_3^2 = 2 \pi \sqrt{\alpha'}/ (R_3 \ell_6)$ is the
three-dimensional gauge coupling \footnote{In this configuration (without \D5-branes) there is also an extra contribution coming from the \D1s between the $N$-th and the first \D3 brane which gives rise to the $\eta$-superpotential~\cite{Davies:1999uw,Katz:1996th} \(   W_\eta = \eta \, e^{\Sigma_{N}-\Sigma_1}\) as discussed in~\cite{Amariti:2015yea}. %
This will in general not play a role for our monopole quivers where \(\eta \) amounts to a field redefinition.},
. 
Similarly, a Euclidean fundamental string stretched between the \NS5 and the \D3 branes will give a contribution
\begin{equation}
  e^{-S_{\F1}} = \exp[- \frac{\sqrt{\alpha'}}{R_3} \frac{\Delta \Sigma}{e_3^2}] ,
\end{equation}
which vanishes in the \(R_3 \to 0\)  limit that we take.

\paragraph{Monopole quivers.}

We can avoid the breaking of the gauge group to $U(1)^{N}$ due to the mutual repulsion of the \D3 branes 
by attaching them to \D5 branes. 
We will consider only the stable case $\# \D5 > \# \D3$ as we are interested in the dimensional reduction of Seiberg duality.
In the \tIIA frame, we start with $F$ \D6--branes extended along $0123789$ and sitting on the \NS'--brane. The \D6 branes become \D5--branes after T--duality. The strings stretched between the stack of \D3--branes and the \D5s correspond to 
$F$ massless fundamentals $Q$ and anti-fundamentals $\widetilde Q$.

When \D5--branes sitting at $x_3 =0$ intersect the worldsheet of the \D1--strings,
they contribute two additional zero modes to the \D1--instanton and the superpotential in Eq.~(\ref{WD1}) is not generated.
So the \D5--branes have the effect of screening the repulsive force between the \D3--branes~\cite{deBoer:1997kr,Elitzur:1997hc}.

The effective three-dimensional theories on $S^1$ can undergo non-trivial real mass or Higgs flows. These flows are often necessary in order to recover conventional three-dimensional theories, allowing to preserve dualities when taking the zero-radius limit.
In the brane setup, this corresponds to moving stacks of $n_i$ \D3s and $f_i$ \D5s along the circle, with $f_i > n_i$.
The resulting theory corresponds to a product of $U(n_i)$ gauge factors, each with $f_i$ flavors.
In the \ac{ir}, these gauge sectors do not interact through matter fields.
Considering each sector as decoupled, the $1$-st and the $n_i$-th \D3--brane in each stack are free to move without being subjected to any force, 
so two directions in the moduli space seem to remain unlifted in each sector.
These directions are however lifted by an \ac{ahw} superpotential between the $i$-th and the $i+1$-th gauge sectors. 
Such a superpotential has the form
\begin{equation}
  W = \exp [i \pqty{ \Sigma_i^{(1)} - \Sigma_{i+1}^{(n_{i+1})} }] = T_i \wt T_{i+1},
\end{equation}
where \(\Sigma_i^{(j)}\) refers to the \(j\)-th brane in the \(i\)-th stack.
In gauge theoretical terms, $\Sigma_i^{(1)}$ and $\Sigma_i^{(n_i)}$ are the bare monopoles of the $U(n_i)$ theory with flux $(1,0,\dots,0)$ and $(0,0,\dots,-1)$, respectively.
We will use the notation \(T\) for the former and \(\wt T\) for the latter.

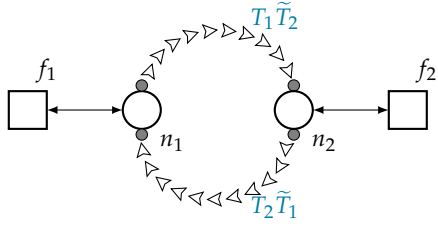
\begin{SCfigure}
  \begin{footnotesize}
    \begin{tikzpicture}[]
  \tikzset{paint/.style={ draw=#1!0!black, fill=#1!50 }, decorate with/.style={decorate,decoration={shape backgrounds,shape=#1,shape size=5pt}}}

  \draw[latex-latex] (-.25,0) -- (-1.25,0);
  \draw[latex-latex] (2.25,0) -- (3.25,0);

  \draw [decorate with=dart, paint=white] (0.1,0.5) arc (165:15:.95 and .75);
  \draw [decorate with=dart, paint=white] (1.9,-0.5) arc (355:185:.95 and .75);

  \draw[fill=white, thick] (0,0) circle [radius=.25] node[label={[label distance=.1cm]-60:\(n_1\)}] {};
  \draw[fill=gray] (0,.325) circle [radius=.075];
  \draw[fill=gray] (2,.325) circle [radius=.075];
  \draw[fill=gray] (0,-.325) circle [radius=.075];
  \draw[fill=gray] (2,-.325) circle [radius=.075];

  \draw[fill=white, thick] (-1.75,-.25) rectangle (-1.25,.25) node[above] {\(f_1\)} ;

  \draw[fill=white, thick] (2,0) circle [radius=.25]  node[label={[label distance=.1cm]-60:\(n_2\)}] {};
  \draw[fill=white, thick] (3.25,-.25) rectangle (3.75,.25) node[above] {\(f_2\)} ;

  \node[cerulean] at (1.75,1.25) {\(T_1 \wt T_2\)};
  \node[cerulean] at (1.75,-1.25) {\(T_2 \wt T_1\)};

\end{tikzpicture}
  \end{footnotesize}
  \caption{A \emph{monopole quiver}. The gauge groups \(U(n_1)\) and \(U(n_2)\) are represented as circles, the flavors as squares and the monopole interactions as darts joining the first brane in the first group to the last in the second group and the first brane in the second group to the last one in the first group.}
  \label{fig:monopole-quiver}
\end{SCfigure}

The resulting theory is thus akin to a quiver gauge theory in which each of the stacks of \D3s corresponds to a node in the quiver.
When we take the (dual) circle radius to be very large, the bifundamental fields corresponding to fundamental strings stretched between the stacks vanish, while the monopoles corresponding to the \D1 branes stretched between the stacks of \D3s remain, due to their different \(R_3\)-dependence in the action, as pointed out above.
The stacks of \D3s thus only interact via the monopoles, which is why we refer to the resulting quiver as \emph{monopole quiver}. 
The simplest case, with two gauge groups and superpotential
\begin{equation}
  W = T_1 \wt T_2 + \wt T_1 T_2,
\end{equation}
is represented in Fig.~\ref{fig:monopole-quiver}.
Using this set-up as a starting point, we can perform a variety of duality transformations via \ac{hw} transitions just as in~\cite{Amariti:2015yea,Amariti:2015mva,Amariti:2016kat}, resulting in new dualities. 

\paragraph{Adding orientifolds.}

Adding orientifold planes to our brane set-up allows us to construct theories with real orthogonal and symplectic gauge groups
and matter fields in the symmetric and antisymmetric representation of the gauge group, see \emph{e.g.} the reviews~\cite{Giveon:1998sr,Amariti:2016kat}.
An orientifold is the combined action of a parity inversion $\sigma$ of the coordinates transverse to the
plane, a world-sheet parity $\Omega$ and $(-1)^{F_L}$, $F_L$ being the left-moving fermion number.
There are \(p\)-dimensional orientifold planes (\O{p} planes) with even $p$ in \tIIA and odd $p$ in \tIIB.
We will study the case of $p=4$ in \tIIA.
There are in total four possibilities, $Op^{\pm}$ and $\widetilde{Op}^{\pm}$~\cite{Hanany:1999sj,Hanany:2000fq}.
Specifying the $\mathbb{Z}_2$ charges characterizes the action of the orientifold on the gauge theory completely.

We will add an \O4 plane on top of the stack of \D4~branes realizing the four-dimensional gauge theory.
It can be shown that 
\begin{itemize}
\item for \(\O4^- \) we get gauge group \(SO(2N)\), 
\item for \(\O4^+\) we have \(Sp(2N)\), 
\item for \(\widetilde{\O4}^-\) we have \(SO(2N+1)\) and 
\item for \(\widetilde{\O4}^+\) we get again \(Sp(2N)\) but with a different
non-perturbative sector. 
\end{itemize}

The three-dimensional system is obtained by compactifying the $x_3$--direction and T--dualizing.
In the presence of a compact direction, orientifold planes come in pairs located at $x_3 = 0$ and at the so-called mirror point $x^{\circ}_3=\frac{\alpha'}{R_3}\pi$
\cite{Hanany:2001iy}.
Here we restrict our attention to the $\O4^{+}$--plane, that will turn into a pair of ($\O3^+,\O3^+$)--planes after T-duality.

\section{A new 3D duality from 4D}\label{sec:newfrom4D}

In this section we derive a new duality by reducing four-dimensional Seiberg duality on $S^1$ and performing a large real mass flow to recover the three-dimensional limit.

The final duality can be summarized as follows:
\begin{itemize}
\item The electric theory is a $U(n_1) \times U(n_2)$ gauge theory with $f_1$ flavors $Q_1$ and $\wt Q_1$ in the first sector and $f_2$ flavors $Q_2$ and $\wt Q_2$ in the second sector, with $f_i > n_i$.
  There is also a superpotential
  \begin{equation}
    \label{Welemon}
    W = \exp[i \pqty{\Sigma_1^{(1)} - \Sigma_2^{(n_2)}}] + \exp[i \pqty{\Sigma_2^{(1)} - \Sigma_1^{(n_1)}}]
    \equiv
    T_1 \wt T_2 + \wt T_1 T_2,
  \end{equation}
  where $\Sigma_1^{(1)}$ and $\Sigma_1^{(n_1)}$ are the bare monopoles of the $U(n_1)$ theory with flux $(1,0,\dots,0)$ and $(0,0,\dots,-1)$, and $\Sigma_2^{(1)}$ and $\Sigma_2^{(n_2)}$ are the bare monopoles of the $U(n_2)$ theory with flux $(1,0,\dots,0)$ and $(0,0,\dots,-1)$.
  The charges of the fields are given in Table~\ref{tab:U-n1-U-n2-charges}.
  \begin{table}
    \begin{tabular}{LRRRRRRR}
      \toprule
                      & U(n_1)         & U(n_2)         & SU(f_1)        & SU(f_2)        & U(1)_A & U(1)_J & U(1)_R                \\
      \midrule
      Q_1             & n_1            & 1              & f_1            & 1              & 1/f_1  & 0      & \Delta_1              \\
      \widetilde{Q}_1 & \overline{n}_1 & 1              & \overline{f}_1 & 1              & 1/f_1  & 0      & \Delta_1              \\
      Q_2             & 1              & n_2            & 1              & f_2            & -1/f_2 & 0      & \Delta_2              \\
      \widetilde{Q}_2   & 1              & \overline{n}_2 & 1              & \overline{f}_2 & -1/f_2 & 0      & \Delta_2              \\
      T_1       & 1              & 1              & 1              & 1              & 1      & 1      & f_1(1-\Delta_1)-n_1+1 \\
      \widetilde{T}_1       & 1              & 1              & 1              & 1              & 1      & -1     & f_1(1-\Delta_1)-n_1+1 \\
      T_2      & 1              & 1              & 1              & 1              & -1     & 1      & f_2(1-\Delta_2)-n_2+1 \\
      \widetilde{T}_2       & 1              & 1              & 1              & 1              & -1     & -1     & f_2(1-\Delta_2)-n_2+1 \\
      \bottomrule
    \end{tabular}
    \caption{Field charges in the electric phase of the duality for the $U(n_1) \times U(n_2)$ monopole quiver.}
    \label{tab:U-n1-U-n2-charges}
  \end{table}
\item The magnetic theory is a $U(f_1-n_1) \times U(f_2-n_2)$ gauge theory with $f_1$ flavors $q_1$ and $\wt q_1$ in the first sector and $f_2$ flavors $q_2$ and $\wt q_2$ in the second sector. There are also the mesons $M_1 = Q_1 \wt Q_1$ and $M_2= Q_2 \wt Q_2$. 
 The superpotential takes the form
  \begin{equation}
    \label{Wmagmon}
    W = M_1 q_1 \wt q_1+M_2 q_2 \wt q_2 + t_1 \wt t_2 + \wt t_1 t_2,
  \end{equation}
  where $t_1$ and $\wt t_1 $ are the bare monopoles of the $U(f_1-n_1)$ theory with flux $(1,0,\dots,0)$ and $(0,0,\dots,-1)$, and $t_2$ and $\wt t_2$ are the bare monopoles of the $U(f_2-n_2)$ theory with flux $(1,0,\dots,0)$ and $(0,0,\dots,-1)$.
\end{itemize}

In the following we will derive the above duality via the reduction of the four-dimensional duality on the brane system and independently via the reduction of the four-\ac{sci} to the three-dimensional partition function on the squashed three-sphere.

\subsection{Derivation from the brane setup}

Consider the reduction of four-dimensional Seiberg duality on $S^1$ in the brane picture~\cite{Giveon:1998sr}. This is done by T-dualizing in the direction $x_3$.
The theory on the circle has one \NS{} and one \NS' brane, with one compact direction. There are $N$ \D3s extended along $x_6$ and $F$ \D5s.
The D--branes are at the origin of the circle.
In order to perform the three-dimensional limit, we select a vacuum in which we move $n_1$ \D3s and $f_1$ \D5s to $x_3=\pi \alpha' f_2/r$ and $n_2$ \D3s and $f_2$ \D5s to $x_3=-\pi \alpha' f_1/r$ on the circle.
The final configuration %
corresponds to a product of two gauge factors interacting via the superpotential in Eq.~(\ref{Welemon}), as can be understood by visualizing the Euclidean \D1 branes discussed above between the two sectors.

The dual phase is obtained via a \ac{hw} transition.
There are $f_1-n_1$ \D3s and $f_1$ \D5s at $x_3=\pi \alpha' f_2/r$ and $f_2-n_2$ \D3s and $f_2$ \D5s at $x_3=-\pi \alpha' f_1/r$ on the circle.
This theory corresponds to a product of two gauge factors interacting through the superpotential (\ref{Wmagmon}).

\subsection{Derivation from the partition function}

Here we show that the duality discussed above can be obtained by 
implementing the real mass flow on the partition function on $S_b^3$. 
The relevant formulas are given in Appendix~\ref{appPF}.
When the integral identity between the four-dimensional Seiberg-dual phases is
reduced on the circle, the partition functions  are related by the  relation
\begin{equation}
\label{initial}
Z_{U(N)} (\Lambda; \mu,\nu) 
= 
\prod_{a,b=1}^{F} \Gamma_h(\mu_a +  \nu_b)
Z_{U(F-N)} (-\Lambda; \omega-\nu,\omega-\mu) 
\end{equation}
with the balancing condition 
\begin{equation}
\sum_{a=1}^{F} \mu_a = \sum_{a=1}^{F} \nu_a = \omega(F-N)
\end{equation}
that signals the presence of a three-dimensional effective duality with an $\eta$-superpotential
in both phases.
In the vacuum chosen above, the real masses split as
\begin{align}
  \mu &\to
  \begin{cases} 
    m_a^1 + f_2 s \\
    m_a^2 - f_1 s   
  \end{cases} &
                \nu &\to
                \begin{cases}
                  n_a^1 - f_2 s ,& a = 1, \dots, f_1\\
                  n_a^2 + f_1 s , & a = 1, \dots, f_2.
                \end{cases}
\end{align}
The gauge group is broken by the choice of the vacuum
\begin{equation}
  \sigma \rightarrow
  \begin{cases}
    \sigma_i^1 - f_2 s , & i=1,\dots,n_1 \\
    \sigma_i^2 + f_1 s , & i=1,\dots,n_2.
  \end{cases}
\end{equation}
The balancing condition becomes
\begin{equation}
  \sum_{a=1}^{f_1} m_a^1 + \sum_{a=1}^{f_2} m_a^2 =
  \sum_{a=1}^{f_1} n_a^1  + \sum_{a=1}^{f_2} n_a^2   = 
  \omega(F-N).
\end{equation}
The real masses can be written in terms of the global symmetries as
\begin{align}
  m_a^1 &= M_a^1 + \frac{m_A}{f_1} + \omega \Delta_1  & 
                                          m_a^2 &= M_a^2  + \frac{m_A}{f_1}  + \omega \Delta_1 \\
  n_a^1 &= N_a^1- \frac{m_A}{f_2} + \omega \Delta_2  &
                                        n_a^2 &= N_a^2  - \frac{m_A}{f_2}  + \omega \Delta_2 
\end{align}
with the constraint on the non-Abelian gauge symmetries
\begin{equation}
  \sum_{a=1}^{f_1} M_a^1 \, = \, \sum_{a=1}^{f_1} N_a^1 \, = \,  \sum_{a=1}^{f_2} M_a^2 \, = \,  \sum_{a=1}^{f_2}  N_a^2  \, = \,  0.
\end{equation}
The balancing condition forces the $R$-charges to be constrained by 
\begin{equation}
  f_1 \pqty{1-\Delta_1} + f_2 \pqty{1-\Delta_2}- n_1-n_2 = 0 .
\end{equation}
The dual theory has real masses inherited from the electric theory and the gauge group is broken as
\begin{equation}
     \widetilde  \sigma \rightarrow
  \begin{cases}
    \widetilde \sigma_i^1 - f_2 s ,& i=1,\dots,f_1-n_1 \\
    \widetilde \sigma_i^2 + f_1 s , & i=1,\dots,f_2-n_2.
  \end{cases}
\end{equation}
At large $s$, there is a divergent phase on both sides of (\ref{initial}).
This divergent term coincides in the electric and in the magnetic phase.
One can ignore it and can compare only the finite terms, arriving to the equality 
\begin{multline}
\label{newd}
  Z_{U(n_1)\times U(n_2)} (\Lambda_1,\Lambda_2; m^1, m^2, n^1,n^2 ) 
  =
  \prod_{a,b}^{f_1} \Gamma_h(m_a^1 + n_a^1) \prod_{a,b}^{f_2} 
  \Gamma_h(m_a^2 + n_a^2) \\
  Z_{U(f_1-n_1)\times U(f_2-n_2)} (-\Lambda_1,-\Lambda_2; \omega-n^1,\omega - n^2, \omega-m^1,\omega- m^2 ) ,
\end{multline}
where the effective \ac{fi} terms are
\begin{align}
  \Lambda_1 &= \Lambda+ 2m_A- 2\omega \pqty{n_2-f_2 \pqty{1-\Delta _2}}, &  \Lambda_2 &= \Lambda+ 2m_A+ 2\omega \pqty{n_1-f_2 \pqty{1-\Delta _1}}.
\end{align}
These \ac{fi} terms are compatible with the presence of the monopole superpotentials in both phases.
The relation (\ref{newd}) corresponds to the new duality between the monopole quivers discussed on the field theory 
side.

\subsection{Relation to Aharony dualities}

\begin{figure}[t]
  \centering
  \begin{footnotesize}
    \begin{tikzpicture}[]
  \tikzset{paint/.style={ draw=#1!0!black, fill=#1!50 }, decorate with/.style=
    {decorate,decoration={shape backgrounds,shape=#1,shape size=5pt}}}
  \begin{scope}[]

    \draw[latex-latex] (-.25,0) -- (-1,0);
    \draw[latex-latex] (2.25,0) -- (3,0);

    \draw [decorate with=dart, paint=white] (0.1,0.5) arc (165:15:.95 and .75);
    \draw [decorate with=dart, paint=white] (1.9,-0.5) arc (355:185:.95 and .75);

    \draw[fill=white, thick] (0,0) circle [radius=.25] node[label={[label distance=.1cm]-60:\(n_1\)}] {};
    \draw[fill=gray] (0,.325) circle [radius=.075];
    \draw[fill=gray] (2,.325) circle [radius=.075];
    \draw[fill=gray] (0,-.325) circle [radius=.075];
    \draw[fill=gray] (2,-.325) circle [radius=.075];

    \draw[fill=white, thick] (-1.5,-.25) rectangle (-1,.25) node[above] {\(f_1\)} ;

    \draw[fill=white, thick] (2,0) circle [radius=.25]  node[label={[label distance=.1cm]-60:\(n_2\)}] {};
    \draw[fill=white, thick] (3,-.25) rectangle (3.5,.25) node[above] {\(f_2\)} ;

    \node[cerulean] at (1.75,1.25) {\(T_1 \wt T_2\)};
    \node[cerulean] at (1.75,-1.25) {\(T_2 \wt T_1\)};
    \node at (1,-2) {(a)};
  \end{scope}

  \begin{scope}[shift={(0,-5)}]

    \draw[latex-latex] (-.25,0) -- (-1,0);
    \draw[latex-latex] (2.25,0) -- (3,0);

    \draw [decorate with=dart, paint=white] (.85,1.1) arc (95:165:.95 and .75);
    \draw[fill=gray] (1,1.1) circle [radius=.075];
    \draw [decorate with=dart, paint=white] (1.15,1.1) arc (85:5:.95 and .75);

    \draw [decorate with=dart, paint=white] (0.1,-.5) arc (185:265:.95 and .75);
    \draw[fill=gray] (1,-1.2) circle [radius=.075];
    \draw [decorate with=dart, paint=white] (1.9,-0.5) arc (355:275:.95 and .75);

    \draw[fill=white, thick] (0,0) circle [radius=.25]  node[label={[label distance=.1cm]-60:\(f_1 - n_1\)}] {};
    \draw[fill=gray] (0,.325) circle [radius=.075];
    \draw[fill=gray] (2,.325) circle [radius=.075];
    \draw[fill=gray] (0,-.325) circle [radius=.075];
    \draw[fill=gray] (2,-.325) circle [radius=.075];

    \draw[fill=white, thick] (-1.5,-.25) rectangle (-1,.25)  node[above] {\(f_1\)} ;
    \draw[-latex] (-1.5,.25) arc (30:330:.5);
    
    \draw[fill=white, thick] (2,0) circle [radius=.25]  node[label={[label distance=.1cm]-60:\(n_2\)}] {};
    \draw[fill=white, thick] (3,-.25) rectangle (3.5,.25)  node[above] {\(f_2\)} ;

    \node[cerulean] at (-.4,1) {\(T_1 \wt t_1\)};
    \node[cerulean] at (2.4,1) {\(T_1 \wt T_2\)};
    \node[cerulean] at (-.4,-.85) {\(t_1 \wt T_1\)};
    \node[cerulean] at (2.4,-.85) {\(T_2 \wt T_1\)};

    \node at (1,-2) {(c)};

  \end{scope}

  \begin{scope}[shift={(7,0)}]

    \draw[latex-latex] (-.25,0) -- (-1,0);
    \draw[latex-latex] (2.25,0) -- (3,0);

    \draw [decorate with=dart, paint=white] (0,.5) arc (165:95:.95 and .75);
    \draw[fill=gray] (1,1.1) circle [radius=.075];
    \draw [decorate with=dart, paint=white] (2,.5) arc (15:85:.95 and .75);

    \draw [decorate with=dart, paint=white] (0.8,-1.2) arc (260:185:.95 and .75);
    \draw[fill=gray] (1,-1.2) circle [radius=.075];
    \draw [decorate with=dart, paint=white] (1.2,-1.2) arc (280:355:.95 and .75);

    \draw[fill=white, thick] (0,0) circle [radius=.25]  node[label={[label distance=.1cm]-60:\(n_1\)}] {};  
    \draw[fill=gray] (0,.325) circle [radius=.075];
    \draw[fill=gray] (2,.325) circle [radius=.075];
    \draw[fill=gray] (0,-.325) circle [radius=.075];
    \draw[fill=gray] (2,-.325) circle [radius=.075];

    \draw[fill=white, thick] (-1.5,-.25) rectangle (-1,.25)  node[above] {\(f_1\)} ;
    \draw[-latex] (3.5,.25) arc (150:-150:.5);

    \draw[fill=white, thick] (2,0) circle [radius=.25]  node[label={[label distance=.1cm]-60:\(f_2 - n_2\)}] {};
    \draw[fill=white, thick] (3,-.25) rectangle (3.5,.25)  node[above] {\(f_2\)} ;

    \node at (1,-2) {(b)};

  \end{scope}

  \begin{scope}[shift={(7,-5)}]

    \draw[latex-latex] (-.25,0) -- (-1,0);
    \draw[latex-latex] (2.25,0) -- (3,0);

    \draw [decorate with=dart, paint=white] (1.9,0.5) arc (15:165:.95 and .75);
    \draw [decorate with=dart, paint=white] (0.05,-0.5) arc (185:355:.95 and .75);

    \draw[fill=white, thick] (0,0) circle [radius=.25]  node[label={[label distance=.1cm]-60:\(f_1 - n_1\)}] {};
    \draw[fill=gray] (0,.325) circle [radius=.075];
    \draw[fill=gray] (2,.325) circle [radius=.075];
    \draw[fill=gray] (0,-.325) circle [radius=.075];
    \draw[fill=gray] (2,-.325) circle [radius=.075];

    \draw[fill=white, thick] (-1.5,-.25) rectangle (-1,.25)  node[above] {\(f_1\)} ;
    \draw[-latex] (-1.5,.25) arc (30:330:.5);

    \draw[fill=white, thick] (2,0) circle [radius=.25]  node[label={[label distance=.1cm]-60:\(f_2 - n_2\)}] {};
    \draw[fill=white, thick] (3,-.25) rectangle (3.5,.25)  node[above] {\(f_2\)} ;
    \draw[-latex] (3.5,.25) arc (150:-150:.5);

    \node[cerulean] at (1.75,1.25) {\(t_1 \wt t_2\)};
    \node[cerulean] at (1.75,-1.25) {\(t_2 \wt t_1\)};

    \node at (1,-2) {(d)};

  \end{scope}

  \node[single arrow,draw,rotate=0, black] at  (4,-1.5) {duality on 2};
  \node[single arrow,draw,rotate=0, black] at  (4,-6.5) {duality on 2};
  \node[single arrow,draw,rotate=-90, black] at  (9.5,-2.5) {duality on 1};%
  \node[single arrow,draw,rotate=-90, black] at  (0,-2.5) {duality on 1};%

\end{tikzpicture}      
  \end{footnotesize}
  \caption{The duality for the \(U(n_1) \times U(n_2)\) monopole quiver can be understood as a sequence of two Aharony dualities. Starting from the electric configuration (a) 
  we can either Aharony-dualize the second node (b) and then the first one, arriving to the configuration (d),
  or Aharony-dualize the first node (c) and then the second one, arriving again to the magnetic configuration (d).
  In the intermediate phases (b) and (c), some monopoles are singlets of the theory.}
  \label{fig:aharony-loop}
\end{figure}
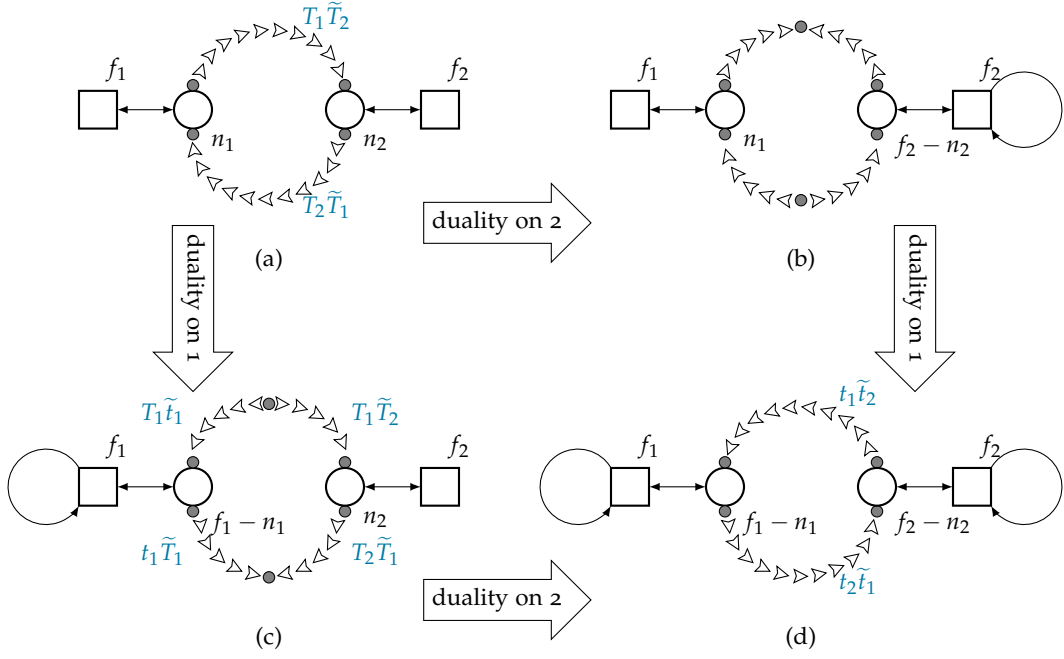

We can also consider the initial configuration and perform a series of Aharony dualities, 
first on the $U(n_1)$ node and then on the $U(n_2)$ node (see Fig.~\ref{fig:aharony-loop})
This gives us a consistency check of the duality discussed above, because 
the final configuration should coincide with the duality derived by a \ac{hw} transition on the brane setup.

Let us study the first duality. The new quiver consists of a $U(f_1-n_1) \times U(n_2)$ 
theory with $f_1$ flavors $q_1$ and $\widetilde q_1$ in the $U(f_1-n_1)$ sector, with a meson $M_1 = Q_1 \wt Q_1$ 
and $f_2$ flavors $Q_2$ and $\widetilde Q_2$ in the $U(n_2)$ sector.
The superpotential of this theory is
\begin{equation}
  W = M_1 q_1 \wt q_1 + T_1 \wt t_1 + \wt T_1 t_1 + T_1 \wt T_2 + \wt T_1 T_2 .
\end{equation}
In this phase, the partition function is given by
\begin{multline}
\Gamma_h \bigg(\pm\frac{\Lambda_1}{2} - \frac{1}{2} \sum_{a=1}^{f_1} (m_a +n_a) + \omega(f_1-n_1+1) \bigg)
  \prod_{a<b} \Gamma_h(m_a+n_a) \\
  \times Z_{U(f_1-n_1) \times U(n_2)} (-\Lambda_1,\Lambda_2; \omega-n_a,\widetilde m_a,\omega-m_a,\widetilde n_a).
\end{multline}
Now we perform the second Aharony duality on the second node.
In this case the theory is the one discussed above, with superpotential 
\begin{equation}
\label{Wfin}
W = M_1 q_1 \wt q_1 +M_2 q_2 \wt q_2 + T_1 \wt t_1 + \wt T_1 t_1 + T_1 \wt T_2 + \wt T_1 T_2 + T_2 \wt t_2 + \wt T_2 t_2,
\end{equation}
where the monopoles $T_i$ and $\wt T_i$ in 
(\ref{Wfin}) should be treated as massive singlets
and integrated out. This leads to the expected dual superpotential 
\begin{equation}
W = M_1 q_1 \wt q_1 +M_2 q_2 \wt q_2 + t_1 \wt t_2 + \wt t_1 t_2.
\end{equation}
We can also reproduce this result on the partition function.
In this case we have
\begin{equation}
  \begin{aligned}
    & \Gamma_h \bigg(\pm\frac{\Lambda_1}{2} - \frac{1}{2}\sum_{a=1}^{f_1} (m_a^1 +n_a^1) + \omega(f_1-n_1+1) \bigg)
    \prod_{a<b} \Gamma_h(m_a^1+n_a^1)  \\
    &\times
    \Gamma_h\bigg(\pm\frac{\Lambda_2}{2} -\frac{1}{2} \sum_{a=1}^{f_2} (m_a^2 +n_a^2) + \omega(f_2-n_2+1)\bigg)
    \prod_{a<b} \Gamma_h( m_a^1 + n_a^1) \\
    &\times
    Z_{U(f_1-n_1) \times U(f_2-n_2)} (-\Lambda_1,-\Lambda_2; \omega-n_a^1,\omega-n_a^2,\omega-m_a^1,\omega- m_a^2).
  \end{aligned}
\end{equation}
The monopoles can be integrated out by imposing the balancing condition and using the relation
$\Gamma_h(\omega\pm x) = 1$. In this way formula (\ref{newd}) is recovered.

Observe that the duality for the monopole quiver discussed here reduces to the ordinary Aharony duality
for $f_2 =1$ and $n_2=0$. In this case, the dual theory corresponds to a $U(f_1-n_1) \times U(1)$ quiver,
where the \ac{sqed} sector has one flavor and it is mirror-dual to the XYZ model.
The fields $Y$ and $Z$, correspond to the monopoles of the $U(1)$ sector, $t_2$ and $\wt t_2$
while the field $X$ is identified with the singlet $q_2 \wt q_2$.
The dual superpotential becomes 
\begin{equation}
  W = M_1 q_1 \wt q_1 +M_2 X + X Y Z + t_{1} Y + \wt t_{1} Z.
\end{equation}
By integrating out the massive singlets $M_2$ and $X$, we end up in the expected Aharony-dual theory.
The monopoles $Y$ and $Z$ act as singlets in the dual theory and they have the same quantum numbers of the
monopoles $T_1$ and $\wt T_1$.

The reduction of the monopole quiver duality to Aharony duality for $f_2=1$ and $n_2=0$ can be obtained also from the 
partition function. In this case one has to use the identity~\cite{Jafferis:2010un}
\begin{equation} 
\label{id}
\int \dd{\sigma} e^{-i \pi \Lambda \sigma }\Gamma_h(\sigma+\omega-\widetilde m) \Gamma_h(-\sigma+\omega-\widetilde n) 
=
\Gamma_h(2 \omega-\widetilde m - \widetilde n) \Gamma_h  \left(\tfrac{\pm \Lambda+ \widetilde m +\widetilde n}{2} \right).
\end{equation}
The first term on the RHS of (\ref{id}) corresponds to the field $X$ and it cancels the contribution of the dual meson $M_1$
in the partition function, as can be seen by using the identity $\Gamma_h(2 \omega-x) \Gamma_h(\omega)=1$.
The second term in (\ref{id})  corresponds to the contribution of the monopoles $Y$ and $Z$,
corresponding to the singlets $T_!$ and $\wt T_1$ in the Aharony duality.
This can be shown with the help of the balancing condition, which corresponds to substituting the relation
\begin{equation}
  \widetilde m +\widetilde n = 2 \omega \pqty{f_1-n_1+1} - \sum_{a=1}^{f_1} \pqty{m_a^1+n_a^1}
\end{equation} 
into (\ref{id}). This gives the monopole contribution in the Aharony-dual phase as expected.
\\
\\
We conclude this section by commenting on the general situation.  One can indeed construct a monopole quiver with 
a generic amount of $K$ $U(n_i)$ gauge factors, by separating the $N$ \D3 branes on the compact $x_3$ direction into $K$ stacks. Stability requires that
$f_i$ \D5 branes are attached to each sector, such that $f_i>n_i$.
The sectors interact through \D1 branes connecting the $n_i$-th and $n_{i+1}$-th sector, realizing a generic monopole
quiver. The dual theory is obtained via a \ac{hw} transition and it generalizes the construction discussed in this section.
One has a set of $U(f_i-n_i)$ \ac{sqcd} sectors, with dual fundamentals interacting with mesons. Again these sectors interact
with each other through \ac{ahw} terms, represented by \D1 branes in this picture.
In this general case, one can check  the matching between the partition functions by engineering the flow on $Z_{S_b^3}$.
We show this derivation in Appendix~\ref{app:generalduality}.

\section{SQCD dualities and orientifolds}
\label{sec:knownD}

In this section we show how to derive \ac{sqcd} dualities from four-dimensional theories with real gauge groups via the brane picture, which in these cases also includes orientifold planes. First we consider the two new dualities that have been derived in~\cite{Benini:2017dud}. In order to reproduce them via a brane set-up, we will use the non-trivial vacuum structure discussed above, this time starting from the four-dimensional
brane representation of the \ac{ip} duality. %
We show that both three-dimensional dualities can be obtained in the brane picture by shifting some D-branes on the T-dual 
circle while sending its radius to infinity.
We conclude the section showing that also the flow leading to the conventional Aharony duality can be engineered from the very same brane picture.

\paragraph{Four dimensions.}
Before starting the three-dimensional analysis, we review the basic aspects of the four-dimensional \ac{ip} duality that will be necessary in the following.

The electric theory has an $Sp(2 N)$ gauge group with $2 F$ fundamentals $Q$ and vanishing superpotential.
The dual theory has $Sp(2 (F-N-2))$  gauge group with $2 F$ fundamentals $q$ and $F(2F-1)$ mesonic operators $M = Q Q$, interacting through the superpotential $W = M q q $. 

In a \tIIA  brane setup, there are $2N$ \D4 branes stretched between one \NS~and one \NS'~brane.
An $\O4^+$ plane on the stack of \D4 branes realizes the projection of $SU(2N)$ to $Sp(2N)$. Flavor is introduced by the addition of $2F$ \D6 branes. 
The brane set-up is summarized in Table~\ref{tab:bbpI}.
The dual theory is obtained via a \ac{hw} transition. In this case, one has $2(F-N-2)$ \D4 branes stretched between the \NS~ and the \NS' brane.%
\footnote{The factor of \(2\) is necessary to preserve the linking number in presence of the $O4^+$  plane.}

\begin{table}
  \centering
  \begin{tabular}{LRRRRRRRRRR}
  \toprule
            & 0 & 1 & 2 & 3 & 4 & 5 & 6 & 7 & 8 & 9 \\
       \midrule
  \D4       & \times & \times & \times & \times &   &   & \times &   &   &   \\
  \D6       & \times & \times & \times & \times &   &   &   & \times & \times & \times \\
  \NS       & \times & \times & \times & \times & \times & \times &   &   &   &   \\
  \NS'      & \times & \times & \times & \times &   &   &   &   & \times & \times \\
  {\O{4}}^+ & \times & \times & \times & \times &   &   & \times &   &   &   \\
  \bottomrule
  \end{tabular}
  \caption{Brane setup realizing the four-dimensional \ac{ip} duality.}
  \label{tab:bbpI}
\end{table}

\paragraph{Reduction.}
The \ac{ip} duality can be reduced on a circle, leading to a duality between an $Sp(2N)$ gauge theory
with $2F$ fundamentals
and an $Sp(2(F-N-2))$ gauge theory with $2F$ fundamentals interacting with $F(2F-1)$ mesons.
The presence of the circle prevents the generation of the axial symmetry by inducing an $\eta$ superpotential 
in both phases.
This is the usual circle reduction of four-dimensional dualities to three dimensions discussed in~\cite{Aharony:2013dha}.

One can flow to three dimensions with a real mass and/or a Higgs flow in a conventional way, obtaining the Aharony duality with symplectic gauge groups, as discussed in~\cite{Amariti:2015mva}.
By performing a different real mass and Higgs flow in the two dual phases, we reproduce the new dualities of~\cite{Benini:2017dud}.
We refer to these dualities as \ac{bbpI} and \ac{bbpII} duality.

\subsection{BBP\textsubscript{I}}

\paragraph{Field theory.} 

The \ac{bbpI} duality is obtained by splitting the $2F$ real masses $\mu_a$ into two sets, $m_a$ and $\widetilde{m}_a$, each with $F$ elements, and performing the shifts 
\begin{align}
  m_a &\rightarrow m_a +s, & n_a &\rightarrow n_a - s.
\end{align}
A Higgs flow $\sigma_i \rightarrow \sigma_i+s$ is taken for both real scalars in the $Sp(2N)$ and in the dual $Sp(2(F-N-2))$ gauge groups.
At large $s$, the massive fields can be integrated out, leading to a three-dimensional duality between two unitary gauge theories. 

\begin{itemize}
\item The electric theory is a three-dimensional $U(N)$ $\mathcal{N}=2$ theory with $F$ fundamental and antifundamental flavors, $Q$ and $\wt Q$.
  There is a superpotential interaction 
  \begin{equation}
    W = T + \wt T,
  \end{equation}
  where $T$ and $\wt T$ are the $(1,0,\dots,0)$ and the $(0,\dots,0,-1)$ monopoles, respectively.
\item The dual theory is a three-dimensional $\mathcal{N}=2$ $U(F-N-2)$  theory with $F$
  fundamental and antifundamental flavors, $q$ and $\wt q$, and $F^2$ singlets $M$.
  There is a superpotential interaction 
  \begin{equation}
    W = M q \wt q +  t + \wt t,
  \end{equation}
  where $t$ and $\wt t$ are the $(1,0,\dots,0)$ and the $(0,\dots,0,-1)$ monopoles, respectively.
\end{itemize}

\paragraph{Brane picture.} The duality just reviewed can be obtained by brane engineering the \ac{ip} duality reduced on the circle.
By compactifying $x_3$ and performing a T-duality, we arrive at a system consisting of an \NS{} brane, an \NS' brane, $2N$ \D3 branes, $2 F$ \D5s and two $O3^+$ planes.
The branes are extended as shown in Table~\ref{tab:BBPIT}.
\begin{table}
  \centering
  \begin{tabular}{LRRRRRRRRRR}
    \toprule
    & 0 & 1 & 2 & 3 & 4 & 5 & 6 & 7 & 8 & 9 \\
    \midrule
    \D3   & \times & \times & \times &   &   &   & \times &   &   &   \\
    \D5   & \times & \times & \times &   &   &   &   & \times & \times & \times \\
    \NS{} & \times & \times & \times & \times & \times & \times &   &   &   &   \\
    \NS'  & \times & \times & \times & \times &   &   &   &   & \times & \times \\
    \O3^+ & \times & \times & \times &   &   &   & \times &   &   &   \\
    \bottomrule
  \end{tabular}
  \caption{Brane setup realizing the reduction of \ac{ip} duality to three dimensions.}
  \label{tab:BBPIT}
\end{table}

The $2N$ \D3s, the $2F$ \D5s and one $\O3^+$ are initially placed at the origin of $x_3$. 
The second $\O3^+$ is at the point $x_3^\circ = \pi \alpha'/R_3$.
Now we shift the branes along \(x^3\): the configuration of interest has  $\pqty{F-2}$ \D5s and $N$  \D3 branes moved to $x_3 = \pi \alpha'/(2R_3)$, and symmetrically $\pqty{F-2}$ \D5s and $N$  \D3 branes moved to $x_3 = 3\pi \alpha'/(2R_3)  =  -\pi \alpha'/(2R_3) $.
Two \D5s are left at $x_3=0$ and the last two are at $x_3^\circ = \pi \alpha'/R_3$.

After a \ac{hw} transition we obtain $F-2$ \D5s and the $F-N-2$  \D3 branes at $x_3 = \pi \alpha'/(2R_3)$, $F-2$ \D5 and the $\pqty{F-N-2}$  \D3 branes at $x_3  =  -\pi \alpha'/(2R_3) $.
There are again two \D5s at $x_3=0$ and two are at $x_3 = \pi \alpha'/R_3$.

\begin{figure}
  \begin{tikzpicture}
  \begin{scope}
    \draw (0,0) circle (2cm);
    \node at (0,0) {(a)};
    \fill (-2,0) circle [radius=2pt];
    \fill (2,0) circle [radius=2pt];
    \node [align=center] at (-2.75,0) {\(\O3^+\) \\ \(2 N\) \D3 \\ \(2F\) \D5};
    \node at (2.5,0) {\(\O3^+\)};
    \node at (-1.25,0) {\(x_3 = 0\)};
    \node at (1.6,0) {\( x_3^\circ\)};
  \end{scope}

  \begin{scope}[shift={(7.5,0)}]
    \draw (0,0) circle (2cm);
    \node at (0,0) {(b)};
    \fill (-2,0) circle [radius=2pt];
    \fill (2,0) circle [radius=2pt];
    \fill (0,2) circle [radius=2pt];
    \fill (0,-2) circle [radius=2pt];
    \node [align=center] at (-2.5,0) {\(\O3^+\) \\ \(2 \) \D5};
    \node [align=center] at (2.5,0) {\(\O3^+\) \\ \(2\) \D5};
    \node [align=center] at (0,2.5) {\(N\) \D3 \\ \(\pqty{F - 2} \) \D5};
    \node at (0,1.5) {\(x_3 = \tfrac{ \alpha' \pi}{2 R_3} \)};
    \node [align=center] at (0,-2.75) {\(N\) \D3 \\ \(\pqty{F - 2} \) \D5};
    \node at (0,-1.5) {\(x_3 = - \tfrac{ \alpha' \pi}{2 R_3} \)};
  \end{scope}

  \begin{scope}[shift={(0,-8)}]
    \draw (0,0) circle (2cm);
    \node at (0,0) {(d)};
    \fill (-2,0) circle [radius=2pt];
    \fill (2,0) circle [radius=2pt];
    \fill (0,2) circle [radius=2pt];
    \fill (0,-2) circle [radius=2pt];
    \node at (-2.5,0) {\(\O3^+\)};
    \node at (2.5,0) {\(\O3^+\)};
    \node [align=center] at (0,2.5) {\(\pqty{F - N - 2}\) \D3 \\ \(F \) \D5};
    \node [align=center] at (0,-2.75) {\(\pqty{F - N - 2}\) \D3 \\ \(F \) \D5};
  \end{scope}

  \begin{scope}[shift={(7.5,-8)}]
    \draw (0,0) circle (2cm);
    \node at (0,0) {(c)};
    \fill (-2,0) circle [radius=2pt];
    \fill (2,0) circle [radius=2pt];
    \fill (0,2) circle [radius=2pt];
    \fill (0,-2) circle [radius=2pt];
    \node [align=center] at (-2.5,0) {\(\O3^+\) \\ \(2 \) \D5};
    \node [align=center] at (2.5,0) {\(\O3^+\) \\ \(2\) \D5};
    \node [align=center] at (0,2.5) {\(\pqty{F - N - 2}\) \D3 \\ \(\pqty{F - 2} \) \D5};
    \node [align=center] at (0,-2.75) {\(\pqty{F - N - 2}\) \D3 \\ \(\pqty{F - 2} \) \D5};
  \end{scope}

  \node[single arrow,draw,rotate=0, black] at  (3.75,0) {shift};%
  \node[single arrow,draw,rotate=-90, black] at  (7.5,-4) {\ac{hw}};%
  \node[single arrow,draw,shape border rotate=180, black] at  (3.75,-8) {shift};%
  
  \begin{scope}[shift={(10,-10)}]
    \begin{footnotesize}
      \draw[<-] (-0.2,0) arc (-140:135:.4 and .4);
      \draw (0.1,.9) node[]{\(x_3\)};
    \end{footnotesize}
  \end{scope}

\end{tikzpicture}
\caption{Brane content at four points in the direction \(x_3\) for the different phases in the \ac{bbpI} duality.
  The configuration obtained from dimensional reduction is in (a), then the \D3s and some \D5s are moved to realize the electric phase (b); a \ac{hw} transition is performed (c) and finally all the \D5s are moved to the same point to realize the magnetic phase (d). The \D5 branes are projected onto the line.}
  \label{fig:branes-BBPI}
\end{figure}
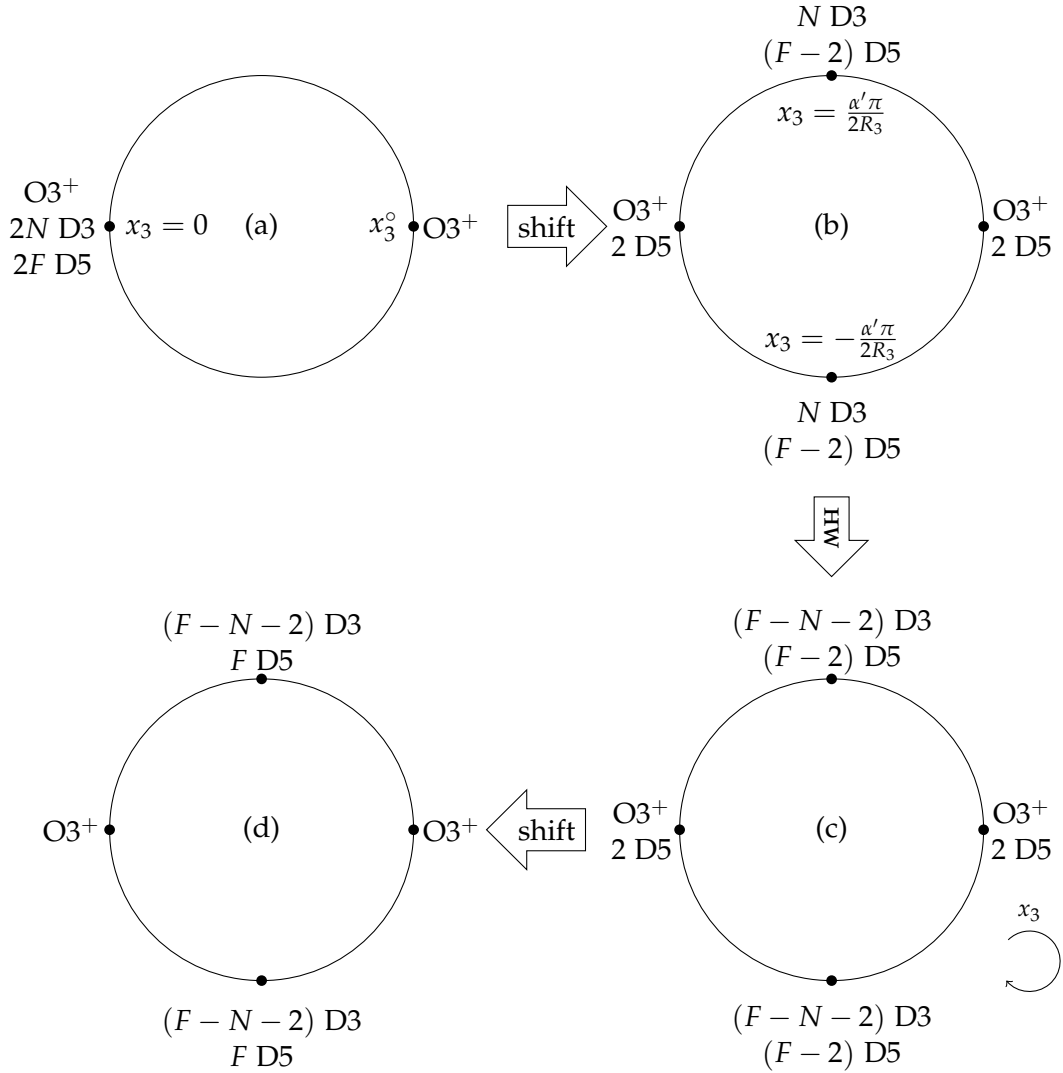

In both phases, we can finally shift the \D5s so that the resulting configuration has $F$ \D5s and the $\pqty{F-N-2}$  \D3 branes at $x_3 = \pi \alpha'/(2 R_3)$ and the same amounts at $x_3 = -\pi \alpha'/(2R_3)$.
A crucial aspect of this construction is that the real mass flow associated to this last shift of the \D5 branes only involves flavor \ac{dof} so that the two theories, originally obtained from a \ac{hw} transition, remain dual at the end of the flow.
The monopole superpotential  can be inferred from the \D1 branes in the usual manner as explained in Sec.~\ref{sec:review}. 
It is $W = T + \wt T$ on the electric side and $W = t + \wt t$ on the magnetic side.
The brane content of the different phases is given in Table~\ref{tab:branes-bbpI} and Figure~\ref{fig:branes-BBPI}.

In the $R_3 \rightarrow 0$  limit, this reproduces the \ac{bbpI} duality.

\begin{table}
  \centering
  \begin{tabular}{lcccc}
    \toprule
                                                & \(x_3 = 0\)     & \(x_3 = \frac{\alpha \pi}{2 R_3}\)       & \(x_3=x_3^\circ \) &\(x_3 = -\frac{\alpha \pi}{2 R_3}\)                                                                                                                                                                                  \\
    \midrule                                                                                                                                                                                               
    \ac{ip} on $S^1$ & \splitcell{c}{\(2N\) \D3}{\(2F\) \D5} & -                                                         & -                              & -                                                         \\
    electric             & \(2\) \D5                             & \splitcell{c}{\(N\) \D3}{\(\pqty{F-2}\) \D5}              & \(2\) \D5                      & \splitcell{c}{\(N\) \D3}{\(\pqty{F-2}\) \D5}              \\
    after \ac{hw}        & \(2\) \D5                             & \splitcell{c}{\(\pqty{F -N -2}\) \D3}{\(\pqty{F-2}\) \D5} & \(2\) \D5                      & \splitcell{c}{\(\pqty{F -N -2}\) \D3}{\(\pqty{F-2}\) \D5} \\
    magnetic             & -                                     & \splitcell{c}{\(\pqty{F -N -2}\) \D3}{\(\pqty{F-2}\) \D5} & -                              & \splitcell{c}{\(\pqty{F -N -2}\) \D3}{\(\pqty{F-2}\) \D5} \\
    \bottomrule                                                                           
  \end{tabular}
  \caption{Brane content at four points in the direction \(x_3\) for the different phases in the \ac{bbpI} duality. In all phases, there is an \(\O3^+\) in \(x_3 = 0\) and \(x_3^\circ \).}
  \label{tab:branes-bbpI}
\end{table}

\subsection{BBP\textsubscript{II}}

\paragraph{Field theory.} 
 The second duality discussed in~\cite{Benini:2017dud} was obtained by considering the \ac{bbpI} model with $\pqty{F+1}$ flavors, shifting the $\pqty{F+1}$-th flavor by 
\begin{align}
  m_{F+1} &\rightarrow m_{F+1} +s , & n_{F+1} &\rightarrow n_{F+1} - s ,
\end{align}
and taking the large-$s$ limit.
\begin{itemize}
\item The electric theory is a three-dimensional $\mathcal{N}=2$  $U(N)$ theory with $F$ fundamental and antifundamental flavors, $Q$ and $\widetilde Q$.
  There is a superpotential interaction,
  \begin{equation}
    \label{eq:WBBPIIele}
    W = T ,
  \end{equation}
  where $T$ is the $(1,0,\dots,0)$ monopole.
\item The dual theory is a three-dimensional $\mathcal{N}=2$ $U(F-N-1)$  theory with $F$
  fundamental and antifundamental flavors, $q$ and $\wt q$, $F^2$ singlets $M$, and a singlet $S$.
  There is a superpotential interaction 
  \begin{equation}
    \label{eq:WBBPIImag}
    W = M q \wt q +  t + S \wt t,
  \end{equation}
  where $t$ and $\wt t$ are the $(1,0,\dots,0)$ and the $(0,\dots,0,-1)$ monopoles.
\end{itemize}

\paragraph{Brane picture.} 

Here we show that the \ac{bbpII} duality can be obtained from the brane setup as well.
Start with the reduction of the \ac{ip} duality on the circle as done above.
In this case, we have $2(F+1)$ \D6 branes and, on the circle, $2N$ \D3s, $2(F+1)$  \D5s plus one $\O3^+$  at the origin $x_3 = 0$ with its dual  $\O3^+$  at   $x_3^\circ = \pi \alpha'/R_3$.

The three-dimensional limit is obtained by shifting $\pqty{F-1}$ \D5s and $N$ \D3 branes to $x_3 = \pi \alpha'/(2R_3)$ and, symmetrically, $\pqty{F-1}$ \D5s and $N$ \D3 branes to $x_3 = 3\pi \alpha'/(2 R_3) = -\pi \alpha'/(2 R_3) $. 
Two \D5s are left at $x_3=0$ and the last two are at $x_3^\circ = \pi \alpha'/R_3$. 
We can make a \ac{hw} transition on this configuration, obtaining $\pqty{F-1}$ \D5s and $\pqty{F-N-1}$ \D3 branes at $x_3 = \pi \alpha'/(2R_3)$ and similarly at $x_3 = -\pi \alpha'/(2 R_3) $.
There are again two \D5s at $x_3=0$ and two at $x_3^\circ = \pi \alpha'/R_3$.

In both phases we can now shift the \D5s to $x_3=0$ (or equivalently the \D5s at $x_3^\circ$) so that the final configuration has $F$ \D5s and all the $\pqty{F-N-1}$ \D3 branes at $x_3 = \pi \alpha'/\pqty{2R_3}$ (and at $x_3 = -\pi \alpha'/\pqty{2R_3}$).
There are also two \D5 branes at $x_3 =\pi \alpha'/R_3 $ (respectively two \D5s at $x_3 =0 $) in both phases.
\begin{itemize}
\item In the electric phase, the three-dimensional limit gives rise directly to the superpotential (\ref{eq:WBBPIIele}), in terms of the \D1 branes connecting the two stacks of \D3 branes 
  through the orientifold at $x_3=0$.
  Furthermore, the absence of massless singlets from the \D5 branes placed at $x_3  =\pi \alpha'/R_3 $ allows us to ignore this sector.
  \item On the magnetic side, the situation is different. We still have the 
  \D1 branes connecting the two stacks of \D3 branes through the orientifold at $x_3=0$ which, like in the electric phase, give a contribution \(W = T\) to the superpotential.
  This time, on the other hand, there is also a massless singlet arising from the \D5 branes placed at $x_3  =\pi \alpha'/R_3 $,
  since the \D5s are parallel to the \NS' brane along the directions $(8,9)$.
  This is the same as the situation discussed in~\cite{Benini:2017dud} for the reduction of the \ac{ip} duality 
  to the Aharony duality for symplectic gauge group. \\
  In the \ac{uv}, \emph{i.e.} for small T-dual radius, the contribution of the \D1 branes from this sector is $W \simeq  t_{high}$.
  Flowing to the \ac{ir}, a scale-matching relation identifies the operator $\widetilde t_{high}$ with the combination $\widetilde{t}_{low} S$, where $S$ is a singlet arising from the \D5 branes which, on the field theory side, is identified with the massless component of the broken meson after the real mass flow.
  All in all, the singlet $S$ interacts with the monopole of the magnetic gauge theory and it has the same quantum number as the electric monopole. 
  The final result coincides with the superpotential (\ref{eq:WBBPIImag}) of the magnetic phase of the \ac{bbpII} duality.
\end{itemize}
The brane content of the different phases is given in Table~\ref{tab:branes-bbpII}.

\begin{table}
  \centering
  \begin{tabular}{lcccc}
    \toprule
                                         & \(x_3 = 0\)     & \(x_3 = \frac{\alpha \pi}{2 R_3}\)       & \(x_3=x_3^\circ \) &\(x_3 = -\frac{\alpha \pi}{2 R_3}\)                                                                                                                                                                                  \\                             
    \midrule                                                                                                                                                                                               
    \ac{ip} on $S^1$ & \splitcell{c}{\(2N\) \D3}{\(2\pqty{F + 1}\) \D5} & -                                                          & -                   & -                                                          \\
    electric             & \(2\) \D5                                        & \splitcell{c}{\(N\) \D3}{\(\pqty{F-1}\) \D5}               & \(2\) \D5           & \splitcell{c}{\(N\) \D3}{\(\pqty{F-1}\) \D5}               \\
    after \ac{hw}        & \(2\) \D5                                        & \splitcell{c}{\(\pqty{F -N - 1}\) \D3}{\(\pqty{F-1}\) \D5} & \(2\) \D5           & \splitcell{c}{\(\pqty{F -N - 1}\) \D3}{\(\pqty{F-1}\) \D5} \\
    magnetic             & -                                                & \splitcell{c}{\(\pqty{F -N - 1 }\) \D3}{\(F\) \D5}         & \(2\) \D5           & \splitcell{c}{\(\pqty{F -N -1}\) \D3}{\(F\) \D5}           \\
    \bottomrule                                                                           
  \end{tabular}
  \caption{Brane content at four points in the direction \(x_3\) for the different phases in the \ac{bbpII} duality. In all phases, there is an \(\O3^+\) in \(x_3 = 0\) and \(x_3^\circ \).}
  \label{tab:branes-bbpII}
\end{table}

\subsection{Aharony duality}

We conclude this section by showing that it is also possible to recover the 
duality of Aharony by reducing the \ac{ip} duality to three dimensions.
The picture is similar to the one discussed above, so we will
be brief in many aspects of the derivation, referring the reader
to the details discussed above when necessary.

\paragraph{Field theory.} 

As already observed in~\cite{Benini:2017dud}, Aharony duality can be obtained from \ac{bbpI} by a real mass deformation.
One starts from $\pqty{F+2}$ flavors and shifts the masses as
\begin{equation}
  \begin{aligned}
    m_{F+1} & \rightarrow m_{F+1} + s, & n_{F+1} &\rightarrow n_{F+1} - s, \\
    m_{F+2} &\rightarrow m_{F+1} -s , & n_{F+2}  &\rightarrow n_{F+2} + s.
  \end{aligned}
\end{equation}
In the large-$s$ limit, the usual Aharony duality is recovered.
The  field content and interactions of Aharony duality can be summarized as follows:
\begin{itemize}
\item The electric theory is a three-dimensional $\mathcal{N}=2$  $U(N)$ theory with $F$ fundamental and antifundamental flavors, $Q$ and $\widetilde Q$
with vanishing superpotential.
\item The dual theory is a three-dimensional $\mathcal{N}=2$ $U(F-N)$  theory with $F$
  fundamental and antifundamental flavors, $q$ and $\wt q$, $F^2$ singlets $M$, and singlets $S$ and $\wt S$.
  There is a superpotential interaction 
  \begin{equation}
    W = M q \wt q + S \wt t + \wt S t ,
  \end{equation}
  where $t$ and $\wt t$ are the $(1,0,\dots,0)$ and the $(0,\dots,0,-1)$ monopoles.
  The singlets $S$ and $\wt S$  are identified respectively with the 
  $(0,\dots,0,-1)$ and the $(1,0,\dots,0,)$ monopoles,
  $\wt T$ and $T$, of the electric theory.
\end{itemize}

\paragraph{Brane picture.} 
Aharony duality can be reproduced at the brane level as above.
In this case we can consider the brane realization of the 
four-dimensional \ac{ip} duality with $2(F+2)$ \D6 branes plus $2N$ \D3s, $2(F+2)$  \D5s, one $\O3^+$  at the origin $x_3 = 0$ and  a second $O3^+$  at   $x_3^\circ = \pi \alpha'/R_3$.

The three-dimensional limit is obtained by shifting $F$ \D5s and $N$  \D3 branes 
to $x_3 = \pi \alpha'/(2R_3)$, and symmetrically $F$ \D5s and  $N$  \D3 branes to $x_3 = 3\pi \alpha'/(2R_3)  =  -\pi \alpha'/(2R_3) $, and we have again two \D5s at $x_3=0$ and two at $x_3^\circ = \pi \alpha'/R_3$.

We can make a \ac{hw} transition on this configuration obtaining $F$ \D5s and the $F-N$  \D3 branes at $x_3 = \pi \alpha'/(2R_3)$, and similarly at $x_3  =  -\pi \alpha'/(2R_3) $.

The final step consists in implementing the large-$s$ limit, considering the massless modes in the spectrum that survive because of the mesons, as discussed in the derivation of the \ac{bbpII} duality.
The brane content of the different phases is given in Table~\ref{tab:branes-Aharony}.

\begin{table}
  \centering
  \begin{tabular}{lcccc}
    \toprule
                                  & \(x_3 = 0\)     & \(x_3 = \frac{\alpha \pi}{2 R_3}\)       & \(x_3=x_3^\circ \) &\(x_3 = -\frac{\alpha \pi}{2 R_3}\)                                                                                                                                                                                  \\
    \midrule                                                                                                                                                                                               
    \ac{ip} on $S^1$ & \splitcell{c}{\(2N\) \D3}{\(2\pqty{F + 2}\) \D5} & -                                                  & -                   & -                                                \\
    electric             & \(2\) \D5                                        & \splitcell{c}{\(N\) \D3}{\(F\) \D5}                & \(2\) \D5           & \splitcell{c}{\(N\) \D3}{\(F\) \D5}              \\
    after \ac{hw}        & \(2\) \D5                                        & \splitcell{c}{\(\pqty{F -N }\) \D3}{\(F\) \D5}     & \(2\) \D5           & \splitcell{c}{\(\pqty{F -N}\) \D3}{\(F\) \D5}    \\
    \bottomrule                                                                           
  \end{tabular}
  \caption{Brane content at four points in the direction \(x_3\) for the different phases in the Aharony duality. In all phases, there is an \(\O3^+\) in \(x_3 = 0\) and \(x_3^\circ \).}
  \label{tab:branes-Aharony}
\end{table}

In this case there are two mesonic sectors in which extra massless matter emerges, leading to the monopole superpotential in the dual phase.
This can be visualized as follows.
At large $s$, the electric theory is $U(N)$ \ac{sqcd} with $F$ flavors.

On the magnetic side, the situation is analogous to the one described above. We have a $U(F-N)$ \ac{sqcd} with $F$ dual flavors interacting with a meson $M$.
There are also extra singlets, arising from the original mesonic operators, because the \D5s at  $x_3 = 0$ and  $x_3^\circ$
are parallel to the \NS' branes along the directions $(8,9)$.
The monopole superpotential can be reconstructed as follows.
One starts by placing the \D1 branes in the \ac{uv} description, \emph{i.e.} when the T-dual radius is small.
They give rise to a \ac{uv} monopole superpotential on both the electric and the magnetic side,
 $W_{UV}^{(ele)}= T_{high} + \widetilde T_{high}$ and 
 ${W}_{UV}^{(mag)}= t_{high} + \widetilde t_{high}$.
The flow to the \ac{ir} is done by sending the T-dual radius to be large.
In the electric theory, in  absence of massless singlets at  $x_3 = 0$ and  $x_3^\circ$, we can safely remove the monopole superpotential.
In the dual phase, one has to consider a scale-matching relation for the two monopoles $t$ and $\tilde t$, in terms the massless singlets arising from the \D5 branes.
They correspond, on the field theory side, to some massless components of the original meson $M$.
By looking at the charge structure, the scaling can be formulated as
\begin{align}
  t_{high} &= t_{low} \wt S, &
                                      \widetilde t_{high} &= \widetilde t_{low} S .
\end{align}
The singlets appearing in these relations have the same charges as the electric monopoles  $T$ and $\wt T$.
By substituting the rescaled monopole into the superpotential $W_{UV}^{(mag)}$, one recovers the expected results for Aharony duality.

\section{Monopole quiver dualities with orientifolds}\label{sec:general}

The monopole quivers considered in Section~\ref{sec:newfrom4D} can be also be constructed in the presence of 
orientifolds.
By considering the reduction of four-dimensional theories on $S^1$ one ends up with pairs of $O3$ planes at
$x_3=0$ and $x_3=x_3^{\circ}$, as discussed above.
In this case, one can consider non-trivial vacuum structures with stacks of \D3- and \D5 branes
along the compact directions (consistently with the identifications imposed by the orientifolds).
When flowing to the three-dimensional limit, a monopole quiver with both real and unitary gauge groups is generated.
The real groups arise if some stacks of \D3 branes are placed at $x_3=0$ and/or $x_3=x_3^{\circ}$,
while the \D3 branes at different positions in $x_3$ give rise to unitary gauge groups.

In this section, we discuss the simplest realization of such a configuration, a duality involving a monopole quiver with 
$Sp(2n_1) \times U(n_2)$ gauge group, obtained by reducing the four-dimensional \ac{ip} duality on $S^1$ and performing a large mass flow.

\paragraph{Field theory.} Let us consider the flow on the gauge theory side.
In the electric theory we assign the masses to the $2F$  fundamentals as
\begin{equation}
  \mu \rightarrow
  \begin{cases}
    \mu_a & \text{if \(a=1,\dots,2f_1\)}\\
    m_a + s & \text{if \(a=1,\dots,f_2\)}\\
    n_a - s& \text{if \(a=1,\dots,f_2\)}.
  \end{cases}
\end{equation}
We also Higgs the gauge group as
\begin{equation}
  \sigma \rightarrow
  \begin{cases}
    \sigma_i & \text{if \(i=1,\dots,n_1\)}\\
    \wt \sigma_i& \text{if \(i=1,\dots,n_2\)}.
  \end{cases}
\end{equation}
This breaks the gauge theory to an $Sp(2n_1)$ theory with $ 2 f_1$ fundamentals
and a $U(n_2)$ theory with $f_2$ fundamental flavors.
The global symmetry is broken to a non-Abelian part $SU(2f_1) \times SU(f_2)^2$
and an Abelian $U(1)_A \times U(1)_R$. The axial $U(1)_A$ symmetry is a 
combination of the two axial symmetries of the unitary and of the symplectic sector 
and of the topological symmetry arising from shifting the dual photon of 
$U(1) \subset U(n_2)$.
These symmetries are broken to a single $U(1)_A$ by the \ac{ahw} interaction between the monopoles
$Y$ of the $Sp(2n_1)$ sector and the $T$ of the $U(n_2)$ sector.
This interaction is 
\begin{equation}
  W = Y \wt T + T .
\end{equation}

The dual theory is obtained by an opportune real mass flow on the $2f_1$ dual fundamentals
and by the Higgsing
\begin{equation}
  \sigma
  \rightarrow
  \begin{cases}
    \sigma_i & i=1,\dots,\wt n_1\\
    \wt \sigma_i& i=1,\dots,\wt n_2,
  \end{cases}
\end{equation}
where $\wt n_1 = f_1 - n_1 -1$ and $\wt n_2 = f_2 - n_2 -1$.
The dual theory becomes an $Sp(2 \wt n_1)$ theory with $ 2 f_1$ fundamentals and a $U(\wt n_2)$ theory with $f_2$ fundamental flavors.
There are also $f_1(2f_1-1)$ mesons $M_1$ in the symplectic sector and $f_2^2$ mesons $M_2$ in the unitary sector.
The superpotential of this dual theory is
\begin{equation}
W = M_1 q_1 q_1 + M_2 q_2 \wt q_2 + y \wt t + t,
\end{equation}
where the monopoles $y$ of the $Sp(2n_1)$ sector and $t, \wt t$ of the $U(n_2)$ sector interact through the \ac{ahw} superpotential leaving the axial $U(1)_A$ symmetry as in the electric theory.

\paragraph{Brane picture.} This duality can be guessed from the brane picture by considering the reduction of the $Sp(2N)$ theory with $2 F$ fundamentals on $S^1$.
After the duality, one can shift $n_2$ \D3s and $\pqty{f_2-1}$ \D5 branes to $x_3 = \pi \alpha'/(2R_3)$, and symmetrically to  $x_3 = -\pi \alpha'/(2R_3)$ plus two \D5 branes at $x_3 = x_3^\circ = \pi \alpha'/r$, thus leaving \(2N - 2n_2 = 2 n_1\) \D3 branes and $2F - 2 f_2 = 2f_1$ \D5 branes at $x_3=0$, on the $O3^+$ plane.
After a \ac{hw} transition we have $\pqty{f_2-n_2-1}$ \D3 and $\pqty{f_2-1}$ \D5 branes at $x_3 = \pm \pi \alpha'/(2R_3)$, \(2\pqty{f_1 - n_1 - 1}\) \D3 and \(2f_1\) \D5 at \(x_3 = 0\) and two \D5 branes at $x_3^\circ$.
Finally, we are free to move one \D5 placed at $x_3^\circ$  to $x_3 = \pi \alpha'/(2R_3)$ and the other one to $x_3 = -\pi \alpha'/(2R_3)$, to recover the magnetic phase.
This motion preserves the duality, being just a real mass flow on the flavor side. 
By placing the \D1 branes between the stacks of \D3 branes we can also read the monopole superpotential in both phases.

The various configurations are collected in Tab.~\ref{tab:branes-SpU}.

\begin{table}
  \centering
  \begin{tabular}{lcccc}
    \toprule
                            & \(x_3 = 0\)     & \(x_3 = \frac{\alpha \pi}{2 R_3}\)       & \(x_3=x_3^\circ \) &\(x_3 = -\frac{\alpha \pi}{2 R_3}\)                                                                                                                                                                                  \\
    \midrule                                                                                                                                                                                               
    \ac{ip} on $S^1$     & \splitcell{c}{\(2N\) \D3}{\(2 F\) \D5}           & -                                                                   & -         & -                                                                   \\
    electric                & \splitcell{c}{\(2n_1\) \D3}{\(2f_1\) \D5}                  & \splitcell{c}{\(n_2\) \D3}{\(\pqty{f_2 - 1}\) \D5}                  & \(2\) \D5 & \splitcell{c}{\(n_2\) \D3}{\(\pqty{f_2 - 1}\) \D5}                  \\
    after \ac{hw}                 & \splitcell{c}{\(2\pqty{f_1 - n_1 - 1}\) \D3}{\(2f_1\) \D5} & \splitcell{c}{\(\pqty{f_2 - n_2 - 1}\) \D3}{\(\pqty{f_2 - 1}\) \D5} & \(2\) \D5 & \splitcell{c}{\(\pqty{f_2 - n_2 - 1}\) \D3}{\(\pqty{f_2 - 1}\) \D5} \\
    magnetic                & \splitcell{c}{\(2\pqty{f_1 - n_1 - 1}\) \D3}{\(2f_1\) \D5} & \splitcell{c}{\(\pqty{f_2 - n_2 - 1}\) \D3}{\(f_2\) \D5}            & -         & \splitcell{c}{\(\pqty{f_2 - n_2 - 1}\) \D3}{\(f_2\) \D5}  \\
    \bottomrule                                                                           
  \end{tabular}
  \caption{Brane content at four points in the direction \(x_3\) for the different phases in the duality for the \(Sp(n_1) \times U(n_2) \) quiver. In all phases, there is an \(\O3^+\) at \(x_3 = 0\) and at \(x_3^\circ \).}
  \label{tab:branes-SpU}
\end{table}

\paragraph{Partition function.} We can corroborate the duality by performing the real mass flow on the partition function.
In this case we use trick of~\cite{Benini:2017dud}, based on 
the symmetry of the integrals.
The large-$s$ behavior cancels in the two theories, because the $s$-dependent phase is
\begin{equation}
-4 s \left(\omega  \left(2 f_2 n_1+n_2^2+n_2\right)-n_1 \sum _{a=1}^{f_2} \left(m_a+n_a\right)\right).
\end{equation}
After eliminating this phase we obtain the equality 
\begin{multline}
  \int \prod_{i=1}^{n_1} \dd{\sigma_i} \prod_{i=1}^{n_2} \dd{\widetilde \sigma_i }
  e^{\lambda_e \widetilde \sigma_i}
  \frac{
    \prod_{i=1}^{n_1}  \prod_{a=1}^{2 f_1} \Gamma_h (\pm \sigma_i + \mu_a)
    \prod_{i=1}^{n_2}  \prod_{a=1}^{f_2} \Gamma_h ( \widetilde \sigma_i + m_a)
    \Gamma_h (- \widetilde \sigma_i + n_a)}
  {
    \prod_{i<j}^{n_1} \Gamma_h(\pm \sigma_i \pm \sigma_j)
    \prod_{i=1}^{n_1} \Gamma_h(\pm 2 \sigma_i )
    \prod_{i<j}^{n_2} \Gamma_h(\pm ( \widetilde \sigma_i -  \widetilde  \sigma_j))}
  \\
  =e^{\phi_m} \prod_{a<b}^{2f_1} \Gamma_h (\mu_a + \mu_b)
  \prod_{a,b=1}^{f_2} \Gamma_h (m_a + n_b)
  \int \prod_{i=1}^{\widetilde n_1} \dd{ \sigma_i} \prod_{i=1}^{\widetilde n_2} \dd{\widetilde \sigma_i }
  e^{\lambda_m \widetilde \sigma_i}
  \\
  \frac{
    \prod_{i=1}^{\widetilde n_1}  \prod_{a=1}^{2 f_1} \Gamma_h (\pm \sigma_i +\omega-\mu_a)
    \prod_{i=1}^{ \widetilde n_2}  \prod_{a=1}^{f_2} \Gamma_h ( \widetilde \sigma_i +\omega- n_a)
    \Gamma_h (- \widetilde \sigma_i + \omega- m_a)
  }{
    \prod_{i<j}^{\widetilde  n_1} \Gamma_h(\pm \sigma_i \pm \sigma_j)
    \prod_{i=1}^{\widetilde  n_1} \Gamma_h(\pm 2 \sigma_i )
    \prod_{i<j}^{\widetilde  n_2} \Gamma_h(\pm ( \widetilde \sigma_i -  \widetilde  \sigma_j))
  }
\end{multline}
with the balancing condition
\begin{equation}
  \sum_{a=1}^{2 f_1} \mu_a + \sum_{a=1}^{f_2} (m_a + n_a) = 2 \omega (F-N -1) \, .
\end{equation}
The phase $\phi_m$ and the \ac{fi} terms are given by
\begin{equation*}
  \begin{aligned}
    \phi_m &=
    \sum _{a=1}^{f_2} (m_a- n_a) \left(\sum _{a=1}^{f_2} (m_a+n_a)-2 \omega  (f_2-3 n_2-1) \right)-(2 n_2+1) \sum _{a=1}^{f_2}(m_a^2-
   n_a^2), \\
    \lambda_e &= 2 \sum _{a=1}^{f_2} (m_a+n_a)-4 \omega  \left(f_2-n_2-1\right), \\
    \lambda_m &= 4 \omega  \left(f_2-n_2\right)-6 \sum _{a=1}^{f_2} \left(m_a+n_a\right) .
  \end{aligned}
\end{equation*}

\section{Conclusions and further developments}\label{sec:concl}

In this paper we have discussed large classes of three-dimensional dualities for $\mathcal{N}=2$ quivers, defined as sets of decoupled \ac{sqcd} sectors interacting through \ac{ahw} interactions.
These dualities are obtained by reducing four-dimensional Seiberg-dual theories on a circle and triggering real mass and Higgs flows. 

We have derived the dualities from the brane-engineering of the gauge theories and corroborated our results via the matching of the three-dimensional partition functions, computed from the circle reduction of the four-dimensional superconformal index. 
We have also shown that in presence of orientifolds, our brane construction allows to recover the dualities recently discovered in~\cite{Benini:2017dud} from the reduction of the \acl{ip} duality.

\medskip
The \ac{ip} duality is realized at the brane level by adding an $\O4^{+}$ plane to the usual setup of \ac{sqcd}.  
After T-duality it turns into a pair of $(\O3^+,\O3^+)$ planes on $S^1$.
When considering pairs of $\O3$ planes on $S^1$, there are in general six possibilities.
Three of them correspond to the compactification of four-dimensional theories with $SP(2N)$, $SO(2N+1)$ or $SO(2N)$ gauge groups, while the others correspond to twisted compactifications with an outer automorphism.
It would be interesting to study the reduction of these theories and the flow in the non-trivial vacua of the type
discussed here.
However, an immediate problem arises when studying the reduction of the index to the partition function for four-dimensional dualities with orthogonal gauge groups.
This procedure is well-defined on the field theory side and on the brane side but the reduction of the index on the circle produces a divergent partition function~\cite{Aharony:2013kma}.
We expect that a double-scaling limit would have to be performed in order to recover the three-dimensional limit discussed here.
In the orthogonal case it would be also interesting to study the reduction of theories with different global 
properties~\cite{Aharony:2013hda,Aharony:2013kma} and to study implications for unitary theories obtained after the three-dimensional real mass flow.

One can also study the reduction of four-dimensional theories with tensor matter. These have a known D-brane realization and the identities between the superconformal index of the dual phases have been listed in~\cite{Spiridonov:2009za,Spiridonov:2011hf}.
It should be possible, for example, to study the reduction of the duality of~\cite{Intriligator:1995ax} for $Sp(2N)$ gauge theories with tensor matter and trigger a flow to the duality of~\cite{Kim:2013cma} for three-dimensional $U(N)$ \ac{sqcd} with adjoint matter.
Another important aspect that we did not study here regards the $U(N)$ dualities with higher powers in the monopole superpotential, discussed in~\cite{Benini:2017dud}.
In the brane picture this should be related to multiple stacks of \D1 branes extended along the \D3- and the \NS{}-branes along the directions $x_3$ and $x_6$.
In would be interesting to check this guess and study the \ac{hw} transition in presence of such an effect. 
It would be also interesting to test the dualities proposed here by matching other partition functions obtained through localization on different compact manifolds, as for example the three-dimensional superconformal index~\cite{Kim:2009wb} and the topologically twisted index of~\cite{Benini:2015noa}.

\bigskip
In general, we have proposed new dualities without focusing on the possible presence of accidental symmetries
in the \ac{ir}~\cite{Morita:2011cs,Agarwal:2012wd,Safdi:2012re,Lee:2016zud}. 
They can arise from the presence of gauge singlets with scaling dimension $\Delta$ below the unitarity bound, $\Delta=\frac{1}{2}$.
This possibility can be checked by maximizing the free energy obtained from the $S^3$ partition function in terms of the  $R$-charges. 
A complete understanding would require finding a \ac{uv} completion of our models in the spirit of~\cite{Benini:2017dud}, which deserves further investigation but is beyond the scope of our present analysis.

We wish to conclude our discussion with a comment on the real mass and Higgs flows from the brane construction.
As already observed in~\cite{Hanany:2001iy}, the displacement of the branes to generic points on the circle requires 
switching on a Wilson line in the gauge theory.
This can be understood in the brane picture in terms of repulsive interactions between \D3 branes.
One realization of this phenomenon is obtained if we place our brane construction into a curved background that preserves the right symmetries and provides the (unique) Wilson line parameter for the quiver on the circle.
This is role is played by the fluxtrap background introduced and studied in~\cite{Hellerman:2011mv,Hellerman:2012zf}.

\section*{Acknowledgments}
A.A.~is grateful to the Galileo Galilei Institute for Theoretical Physics (\textsc{ggi}) for hospitality and 
\textsc{infn} for partial support during the completion of this work, within the program 
``New Developments in AdS$_3$/CFT$_2$ Holography''.

The work of A.A.~and S.R.~is supported by the Swiss National Science Foundation (\textsc{snf}) under grant number \textsc{pp}00\textsc{p}2\_157571/1.

\appendix

\section{The general duality for $U(n)$ product groups}
\label{app:generalduality}

In this appendix we study the flow leading to the generic monopole quiver 
duality for $K$ $U(n_I)$ \ac{sqcd} sectors each with $f_I$ pairs of flavors
$Q_I$ and $\widetilde{Q}_I$
and a monopole superpotential coupling the gauge sectors of the form
\begin{equation}
  W = T_1 \wt T_2 + T_2 \wt T_3 + \dots + T_{K} \wt T_1 .
\end{equation}
The dual theory has 
$K$ $U(\widetilde{n}_I = f_I-n_I)$ \ac{sqcd} sectors each with $f_I$ pairs of flavors $q_I$ and $\widetilde{q}_I$,
$f_I^2$ singlets $M_{I}$ for each \ac{sqcd} sector,
and a monopole superpotential coupling the gauge sectors.
In this case, the dual superpotential is
\begin{equation}
W = \sum_{I=1}^{K} M_I q_I \widetilde{q}_I + t_1 \wt t_2 + t_2 \wt t_3 + \dots + t_{K} \wt t_1 .
\end{equation}
This duality can be obtained from the reduction of four-dimensional Seiberg duality for $U(N)$ \ac{sqcd} with $F$ fundamentals on the circle.
After the reduction, one has to perform a real mass flow on the $F$ masses $\mu_a$ and $\nu_a$
and choosing a non-trivial vacuum on the scalars $\sigma_i$ in the vector multiplet, Higgsing the gauge 
theory and reconstructing the quiver.
The \ac{ahw} interactions associated to this Higgsing reconstruct the monopole superpotential.
The dual theory is recovered by performing the opportune real mass flow and Higgsing 
in the dual phase.

In the following we perform the real mass flow and the Higgsing on the equality relating 
the partition functions of the dual phases,
obtained from the reduction of the equality between the \ac{sci} of the 
four-dimensional Seiberg duality.
In this way we automatically obtain the integral identity matching the three-sphere 
partition functions of the proposed duality.

In the electric theory the real masses $\mu_a$ and $\nu_a$  can be shifted in $K$ different sectors
\begin{align}
  \mu_a & \rightarrow m_a^I + s_I , &
                          \nu_a &\rightarrow n_a^I - s_I , &
                                                 a=1,\dots,f_I,
\end{align}
where each $s_I$ is a divergent real contribution.
The scalar $\sigma_i$ can also be shifted in $K$ sectors as
\begin{align}
  \sigma_i &\rightarrow \sigma_i^I - s_I , &
                         i=1,\dots,n_I.
\end{align}
The ranks $f_I$ and $n_I$ and the infinite shifts $s_I$
are constrained by the relations
\begin{align}
  F &= \sum_{I=1}^{K} f_I, &
                          N &= \sum_{I=1}^{K} n_I, &
                                                  \sum_{I=1}^{K} f_I s_I = 0.
\end{align}
The balancing conditions on the real masses $\mu_a$ and $\nu_a$ becomes 
\begin{equation}
  \sum_{I=1}^{K} \sum_{a=1}^{f_I} m_a^I = \sum_{I=1}^{K}  \sum_{a=1}^{f_I} n_a^I = \omega \big(F-N\big).
\end{equation}
In the dual side the real masses 
are shifted accordingly. The dual Higgsing  is
\begin{align}
\widetilde{\sigma}_i &\rightarrow \widetilde{\sigma}_i^I - s_I &
i & =1,\dots,\widetilde{n}_I.
\end{align}
Imposing the real mass and Higgs flows discussed above, the electric and the magnetic partition function
 can be computed at large $s_I$ by integrating out the divergent contributions.
Formally, we are left with a relation
between the electric and the magnetic partition functions 
of the form
\begin{equation}
  \prod_{I=1}^{K} \lim_{s_I \rightarrow \pm \infty} e^{\Phi_e} Z_{U(n_I)}
  = \prod_{I=1}^{K}  \prod_{a,b=1}^{f_I} \Gamma_h(m_a^I + n_b^I)  
  \lim_{s_I \rightarrow \pm \infty} e^{\Phi_m} Z_{U(\widetilde{n}_I)}.
\end{equation}
The electric phase picks up contributions from integrating out the charged matter 
and the vector multiplet.
The magnetic phase has contributions from the charged matter, the meson and the
vector multiplet.
Summarizing, we have
\begin{align}
  \Phi_e &= \Phi_Q + \Phi_V &  \Phi_m &= \Phi_q + \Phi_{\widetilde{V}} + \Phi_M,
\end{align}
where in the electric theory the relevant contributions are
\begin{align}
  & \begin{multlined}
    \Phi_{Q} =   2\sum_{I\neq J} \sign(s_I -s_J) \Bigg[
    \pqty{ 2 \omega f_J -   \sum_{a=1}^{f_J} \pqty{m_a^J+n_a^J}}\sum_{i=1}^{n_I} \sigma_{i}^{I} 
    +  \omega n_I \sum_{a=1}^{f_J} \pqty{m_a^J-n_a^J}  \\
    + \pqty{n_I \sum_{a=1}^{f_J} \pqty{m_a^J + n_a^J}
      - 2 \omega n_I f_J } \pqty{s_I-s_J}
    - \frac{n_I}{2} \sum_{a=1}^{f_J} \pqty{(m_a^J)^2 - (n_a^J)^2} 
    \Bigg]
  \end{multlined} \\
  & \Phi_{V} = -
    2 \omega \sum_{I\neq J} \sign(s_I -s_J) 
    \bqty{2 n_J \sum_{i=1}^{n_I} \sigma_i^I- n_I n_J \pqty{s_I-s_J}},
\end{align}
while in the magnetic theory we have
\begin{align}
  & \begin{multlined}
    \Phi_{q} = \sum_{I\neq J} \sign(s_I -s_J)  \Bigg[
    \widetilde{n}_I \sum_{a=1}^{f_J} \pqty{(m_a^J)^2 - (n_a^J)^2}  +
    2 \sum_{i=1}^{\widetilde{n}_I} \widetilde{\sigma}_i^I \sum_{a=1}^{f_J} \pqty{m_a^J+n_a^J} \\
    - 2 \widetilde{n}_I  \sum_{a=1}^{f_J} (m_a^J+n_a^J) (s_I-s_J)
    \Bigg]
  \end{multlined} \\
  & \Phi_{\widetilde V} = -  2 \omega
    \sum_{I\neq J} \sign(s_I -s_J) 
    \bqty{2 \widetilde{n}_J \sum_{i=1}^{\widetilde{n}_I} \widetilde{\sigma}_i^I- \widetilde{n}_I \widetilde{n}_J \pqty{s_I-s_J}} \\
  & \begin{multlined}
    \Phi_{M} = \sum_{I\neq J} \sign(s_I -s_J) \Bigg[  2 \sum_{a=1}^{f_I} m_a^I \sum_{a=1}^{f_J} n_a^J -
    f_I \sum_{a=1}^{f_J} \pqty{(m_a^J)^2 - (n_a^J)^2} \\
    + 2 \omega f_I  \sum_{a=1}^{f_J} \pqty{m_a^J - n_a^J} - 2 f_I \pqty{\omega f_J- \sum_{a=1}^{f_I}  \pqty{m_a^J + n_a^J} }
    \pqty{s_I-s_J} \Bigg].
  \end{multlined}
\end{align}
The phases can be reorganized into a divergent term, an \ac{fi} term
and a real mass contribution.
The divergent term has to be the same in the electric and in the magnetic phase.
In the electric phase the divergent term is 
\begin{equation}
  2 \sum_{I\neq J} n_I \abs{s_I -s_J} 
  \bqty{\sum_{a=1}^{f_J} \pqty{m_a^J + n_a^J}
    -  \omega \pqty{2f_J-n_J}},
\end{equation}
and it is straightforward to check that it coincides with the one obtained in the magnetic phase.
The electric and the magnetic \ac{fi} terms are 
\begin{align}
  \lambda_e^I \sum_{i=1}^{n_I}   \sigma_{i}^{I}  \equiv 
  & 2 \sum_{i=1}^{n_I}  \sigma_{i}^{I} \sum_{J(\neq I)=1}^K \sign(s_I -s_J) 
     \bqty{2\omega \pqty{f_J-n_J} - \sum_{a=1}^{f_J} \pqty{m_a^J+n_a^J}},
    \\
  \lambda_m^I 
  \sum_{i=1}^{\widetilde{n}_I}\widetilde{\sigma}_i^I \equiv
  - & 2  \sum_{i=1}^{\widetilde{n}_I} \widetilde{\sigma}_i^I \sum_{J(\neq I)=1}^K \sign(s_I -s_J) 
      \bqty{2 \omega (f_J-n_J)
      - \sum_{a=1}^{f_J} (m_a^J+n_a^J)} ,
\end{align}
where in the relations above, $I$ is fixed.
As expected,  $\lambda_e^I=-\lambda_m^I$ for each gauge group $U(n_I)$ in the quiver.
By summing the other real mass contributions to the phase,
most of the terms cancel among the electric and the magnetic theory.
We are left with the contribution
\begin{equation}
  \label{extra}
  2 \sum_{I\neq J} \sign(s_I-s_J) \bqty{ \omega \pqty{f_I-n_I} \sum_{I=1}^{f_J} \pqty{m_a^J-n_a^J} + \sum_{a=1}^{f_I} m_a^I \sum_{a=1}^{f_J} n_a^J}.
\end{equation}
In order to show that it is vanishing as well let us parameterize the real masses as 
\begin{align}
  m_a^I &= \wt m_a^I + m_A^I \frac{ \alpha_I}{f_I}, &
                                                     n_a^I &= \wt n_a^I  + m_A^I \frac{ \alpha_I}{f_I} ,
\end{align}
with the constraints on the non-Abelian symmetries and on the axial symmetries 
\begin{equation}
 \sum_{a=1}^{f_I} M_a^I = \sum_{a=1}^{f_I} N_a^I = 0, \quad I=1,\dots,K;
\quad\quad
\sum_{I=1}^{K} m_A^I \alpha_I = \omega(F-N).
\end{equation}
Observe that the last constraint shows that there are $K-1$ axial symmetries and one is redundant.
The phase (\ref{extra}) becomes 
\begin{equation}
2
\sum_{I\neq J} \sign(s_I-s_J) \alpha_I \alpha_J m_A^I m_A^J = 0.
\end{equation}

\section{Reduction of the superconformal index to 3D}
\label{appPF}
The reduction of four-dimensional dualities as effective dualities on $S_r^1$ and the $r\rightarrow 0$ limit can be, in general, reproduced by localization.
By considering an opportune scaling limit on the fugacities weighting the \ac{sci} $I_{4d}$, it has indeed been shown~\cite{Gadde:2011ia,Dolan:2011rp,Imamura:2011uw,Aharony:2013dha} that
the three-dimensional partition function on a (possibly squashed) three-sphere $Z_{S_b^3}$~\cite{Kapustin:2009kz,Jafferis:2010un,Hama:2010av,Hama:2011ea} 
can be recovered.
The integral identity relating the \ac{sci} of a pair of four-dimensional Seiberg dual phases translates in an identity relating the partition functions of
a pair of effective three-dimensional dualities  on $S_r^1$. The presence of an $\eta$-superpotential, constraining the three-dimensional duality on $S_r^1$, translates
to the reduction of a constraint among the four-dimensional fugacities into a constraint among the three-dimensional real masses. Such constraints are
referred to in the literature as balancing conditions, both in four and in three dimensions. 
These constraints avoid the generation of symmetries in three dimensions which are anomalous in four dimensions (\emph{e.g.} axial symmetries), 
playing essentially the role of the $\eta$-superpotential in localization.

Recovering the usual three-dimensional limit on $Z_{S_b^3}$ requires triggering a real mass and/or a Higgs flow. This corresponds to shifting some (real) parameters in the 
partition function by an infinite amount. In these cases some care is necessary to show that the divergent parts of the identities coincide.
These real mass flows also modify the balancing condition, allowing, in general, the generation of axial-like symmetries, forbidden on $S_r^1$.
In our discussion we made large use of these ideas to corroborate the new three-dimensional dualities obtained from the brane picture. In this appendix we review the
basic formalism, both to make the discussion self-contained and to provide some references for the main formulas used in the body of the paper.

In the following, we briefly summarize some formal aspects of the reduction of $I_4$ to $Z_{S_b^3}$.
Let us start by introducing the notion of the index (see~\cite{Kinney:2005ej,Romelsberger:2005eg,Dolan:2008qi,Spiridonov:2009za} for details). It can be defined as
\begin{equation}
  I = \Tr(-1)^F e^{-\beta H} (pq)^{\frac{\Delta}{2}}
  p^{j_1+j_2-R/2} q^{j_1-j_2-R/2}
  \prod_{a} u_a^{q_a} ,
\end{equation}
where $F$ represents the fermion number, the Hamiltonian $H$ is defined on $S^3 \times \mathbb{R}$,
the fugacities $p$ and $q$ refer to the $SO(4)=SU(2)_l \times SU(2)_r$ isometry of $S^3$, where
 $j_1$ and $j_2$ are the third spin components, and $R$ is the $U(1)$ R-charge.
There are also fugacities $u_a$, referring to the Cartan of the global (and gauge) symmetry group.
The fugacities  $p$ and $q$ satisfy the conditions 
\begin{align}
  \Im(pq) &= 0, & \abs{p/q} &= 1, & \abs{pq} &< 1  .
\end{align}
The \ac{sci} receives non-vanishing contributions only from states satisfying $H = 0$.
The calculation of the \ac{sci} for a gauge theory proceeds by computing the single particle index
and then taking the plethystic exponential. This corresponds in localization to calculating the one-loop determinants 
of the matter and the vector multiplets. They can be formulated as elliptic Gamma functions,
\begin{equation}
\Gamma_e (y;p,q) \equiv \Gamma_e(y) \equiv
\prod_{j,k=0}^{\infty} \frac{1-p^{i+1} q^{j+1}/ y}{1-p^i q^j y},
\end{equation}
where $y$ refers to the fugacities of the global and gauge symmetries.
The gauge-invariant quantities contributing to the index are found by 
integrating over the holonomy of the gauge group.
Finally, one arrives at the formula
\begin{eqnarray}
I_{G} = 
\frac{\kappa^{r_G}}{\abs{W}}
\oint_{T^{r_G}}
\frac{\dd{ z_i}}{2 \pi i z_i}
\prod_{\alpha \in G_+} \Gamma_e^{-1}(z^{\pm \alpha})
\prod_{\rho \in \mathcal{R}_I,\widetilde{\rho} \in \widetilde{\mathcal{R}}_I} \Gamma_e (z^\rho u^{\widetilde \rho} (pq)^{\frac{R_I}{2}}),
\end{eqnarray}
where $\kappa=(p;p)(q;q)$ and $(x;p) = \prod_{k=0}^{\infty} (1-x p^k)$.
Here, the label $\alpha \in G_{+}$ refers to the positive roots of \(G\), and \(\abs{W}\) is the dimension of the Weyl group. The weight $\rho_I$ stands for the representation
of the matter multiplets under the gauge group and the weight $\widetilde{\rho}_I$ refers to the representation
of the matter multiplets under the flavor symmetry group. The fugacity $z$ is in the Cartan of the gauge symmetry and the fugacity $u$ 
is in the Cartan of the flavor symmetry. The integral is over the maximal Abelian torus of $G$, denoted as $T^G$.
The $R$ charge is denoted by $R_I$.

The partition function is obtained by a \ac{kk} reduction on $S^1$ of the states contributing to the four-dimensional
index. The reduction is done on $S_\mathbf{b}^3 \times \tilde S^1$, where $S_\mathbf{b}^3 $ represents
a squashed three-sphere preserving $U(1)^2 \subset SO(4)$ and $b$ is the squashing parameter.
Defining as $\tilde r_1$  the radius of $\tilde S^1$ the fugacities above can be expressed as
\begin{align}
  p &= e^{2 \pi i \tilde r_1 \omega_1}, &
                                   q &= e^{2 \pi i \tilde r_1 \omega_2}, &
                                                                    u_a &= e^{2 \pi i \tilde r_1 \mu_a}, &
                                                                                                       z_i &= e^{2 \pi i \tilde r_1 \sigma_i} .
\end{align}
In the three-dimensional language, $\mu_a$ are real masses for the flavor symmetries and $\sigma_i$ are
the real scalars in the three-dimensional vector multiplets. The parameters $\omega_{1,2}$ are related to the 
squashing parameter $b$ by $\omega_1 = i b$ and $\omega_2 = ib^{-1}$. We also define the linear
combination $\omega \equiv \frac{\omega_1+\omega_2}{2}$.

The four-dimensional superconformal index reduces to the three-dimensional partition function computed on the squashed three-sphere 
$S_\mathbf{b}^3$. The \ac{bps} states contributing to the four-dimensional index have to be \ac{kk} reduced on the circle.
The massless modes in this reduction are the states contributing to the partition function.
In order to perform the \ac{kk} reduction, it is necessary that all the fugacities
appearing in the index flow to unity. 
The limit $\tilde r_1 \rightarrow 0$ corresponds to 
\begin{equation}
  \label{limit43}
  \lim_{
    \tilde r_1 \rightarrow 0}
  \Gamma_e(e^{2\pi i \tilde r_1 x}; e^{2\pi i \tilde r_1 \omega_1}, 
  e^{2 \pi i \tilde r_1 \omega_2})  
  =
  e^{\frac{i \pi^2}{
      6\tilde r_1 \omega_1 \omega_2}
    (x-\omega)}
  \Gamma_h(x; \omega_1, \omega_2) .
\end{equation}

This formula reduces the one-loop determinants of the four-dimensional fields to the ones of the three-dimensional fields, or more
formally, the elliptic gamma functions $\Gamma_e$
to the hyperbolic gamma function $\Gamma_h$
defined as
\begin{equation}
  \label{eq:Gammahvbd}
  \Gamma_h(x;\omega_1,\omega_2) \equiv
  \Gamma_h(x)\equiv 
  e^{
    \frac{i \pi}{2 \omega_1 \omega_2}
    ((x-\omega)^2 - \frac{\omega_1^2+\omega_2^2}{12})}
  \prod_{j=0}^{\infty} 
  \frac
  {1-e^{\frac{2 \pi i}{\omega_1}(\omega_2-x)} e^{\frac{2 \pi i \omega_2 j}{\omega_1}}}
  {1-e^{-\frac{2 \pi i}{\omega_2} x} e^{-\frac{2 \pi i \omega_1 j}{\omega_2}}}.
\end{equation}
Observe that the divergent prefactor in (\ref{limit43}) represents the four-dimensional
gravitational anomalies
and it coincides in the four-dimensional Seiberg-dual phases.
The general expression for the partition function 
of a three-dimensional gauge theory on $S_b^3$ is given by
\begin{equation}
  \label{eq:SquashedSphere}
  Z_{G;k}(\lambda;\vec{\mu}) = \frac{1}{\abs{W}}\int
  \prod_{i=1}^{G} \frac{d \sigma_i}{\sqrt{-\omega_1 \omega_2}}
  e^{\frac{i k \pi \sigma_i^2}{\omega_1 \omega_2}+\frac{2 \pi i \lambda \sigma_i}{\omega_1 \omega_2}}
  \frac{\prod_I \Gamma_h\left(
  \omega \Delta_I + \rho_I(\sigma)+\widetilde \rho_I(\mu)\right)
  }{\prod_{\alpha \in G_+}\Gamma_h\left(\pm\alpha(\sigma)\right)}.
\end{equation}
The integral is performed over the eigenvalues of the scalar $\sigma$ in the Cartan of the gauge group $G$.
The parameter $\mu$ represents the real mass in the Cartan of the flavor symmetry.
The parameter
$\lambda$ is an \ac{fi} term while $k$ refers to the \ac{cs} action, if present in the dynamics (observe that an analogue \ac{cs} term for
the flavor symmetry can be turned on and it is related to the contact terms of the global currents
\cite{Closset:2012vg,Closset:2012vp}).
The $R$ charge is denoted by $\Delta_I$.

\printbibliography

\end{document}